\documentclass[prd,12pt,nofootinbib,preprint]{revtex4}
\usepackage[T1]{fontenc}
\usepackage[utf8x]{inputenc}
\usepackage{amsmath,amssymb}
\usepackage{epsfig}
\usepackage{url}
\usepackage{subfigure}
\usepackage{graphicx}
\usepackage{grffile}
\usepackage[usenames,dvipsnames]{color}
\usepackage{slashed}
\usepackage{array}
\usepackage{multirow}
\usepackage{xcolor}
\usepackage[colorlinks,citecolor=blue]{hyperref}
\usepackage{color}
\usepackage{cancel}
\usepackage{gensymb}
\usepackage{soul}

\def \met{\not \! E_T }

\def\bar {\overline}

\def\be {\begin{equation}}
\def\ee {\end{equation}}
\def\beq {\begin{equation}}
\def\eeq {\end{equation}}
\def\bea {\begin{eqnarray}}
\def\eea {\end{eqnarray}}

\newcommand{\besub}{\begin{subequations}}
\newcommand{\eesub}{\end{subequations}}

\def\beq{\begin{equation}}
\def\eeq{\end{equation}}
\def\barr{\begin{array}}
\def\earr{\end{array}}

\def\q2 {q^2}

\def\bt{\begin{table}}
\def\et{\end{table}}

\def\mET{E_T \hspace{-1.0em}/\;\:}

\begin{document}

\vskip 30pt 

\begin{flushright}
HRI-RECAPP-2022-002
\end{flushright}
\begin{center}
	{\Large \bf Signals for vector-like leptons in an $S_3$-symmetric 2HDM at ILC } \\
		\vspace*{1cm} {\sf ~Indrani Chakraborty$^{a,}$\footnote{indranic@iitk.ac.in},
		            ~Dilip Kumar
                  Ghosh$^{b,}$\footnote{tpdkg@iacs.res.in}, ~Nivedita
                  Ghosh$^{c,}$\footnote{niveditaghosh@hri.res.in}, 
                  ~Santosh Kumar Rai$^{c,}$\footnote{skrai@hri.res.in}}\\
		\vspace{10pt} {\small } { $^a$Department of Physics, Indian Institute of Technology, Kanpur 208 016, India
		     \\$^b$School Of Physical Sciences, Indian Association for the
                  Cultivation of Science,\\ 2A $\&$ 2B, Raja
                  S.C. Mullick Road, Kolkata 700032, India \\ $^c$
                  Regional Centre for Accelerator-based Particle
                  Physics, \\ Harish-Chandra Research Institute,  A CI of Homi Bhabha National
Institute, \\
Chhatnag Road, Jhunsi, Prayagraj 211019, India}
\end{center}

\begin{abstract}
In this work, we explore the signals of an $S_3$-symmetric two Higgs doublet model with two generations of vector-like 
leptons (VLLs) at the proposed International Linear Collider (ILC). The lightest neutral component of the VLL in this model provides a viable dark matter (DM) candidate satisfying the current relic density data as well as circumventing all direct and indirect DM search constraints. 
Some representative benchmark points have been selected with low, medium and high DM masses, satisfying all theoretical and experimental constraints of the model and constraints coming from the DM sector. The VLLs (both neutral and charged) will be produced in pair leading to multi-lepton and multi-jet final states. We show that the ILC will prove to be a much more efficient and useful machine to hunt for such signals compared to LHC. Using traditional cut-based analysis as well as sophisticated multivariate analysis, we perform a detailed analysis of some promising channels containing mono-lepton plus di-jet, di-lepton, four-lepton and four jets along with missing transverse energy in the final state at 1 TeV ILC.

\end{abstract}
\maketitle


\renewcommand{\thesection}{\Roman{section}}  
\setcounter{footnote}{0}  
\renewcommand{\thefootnote}{\arabic{footnote}}  


\section{Introduction}
\label{intro}
The particle spectrum of the Standard Model (SM) being complete with the Higgs boson discovery \cite{Aad:2012tfa,Chatrchyan:2012ufa} still leaves unanswered questions on our understanding of Nature. In a more complete picture being established, 
the SM can be better termed as an effective theory sustainable up to a certain scale. To circumvent the theoretical and experimental shortcomings, several extensions of the SM have been proposed in the literature exploiting new symmetries or
modifications to the space-time with additional spatial dimensions to name a few. In each such extension addition of either 
bosonic or fermionic fields somewhat becomes inevitable. A typical extension which addresses some of the issues in the flavor sector of particle phenomenology involves vector-like fermions with vector-like leptons (VLL) as a natural entity. Unlike the SM fermions, the left- and right-chiral components of VLLs transform identically under the SM gauge symmetry. 
Some beyond SM extensions where such exotic fermions appear naturally are grand unified 
theories \cite{Rosner:2000rd,Das:2016xuc,Joglekar:2016yap,Das:2017fjf}, theories with non-minimal 
supersymmetric extensions \cite{Moroi:1991mg,Moroi:1992zk,Martin:2009bg,Babu:2008ge,Martin:2010dc,Graham:2009gy,Kang:2007ib}, warped or universal extra-dimension \cite{Randall:1999ee,Randall:1999vf,Agashe:2004cp,Agashe:2006wa,Li:2012zy,Huang:2012kz,Biggio:2003kp,Kaplan:2000av,Cheng:1999fw}, 
composite Higgs model \cite{Chivukula:1998wd,Dobrescu:1997nm,He:2001fz,Contino:2006qr,Anastasiou:2009rv,Kong:2011aa,Gillioz:2012se} and little Higgs model \cite{ArkaniHamed:2002qy,Perelstein:2003wd,Carena:2006jx,Matsumoto:2008fq,Han:2003wu}. The phenomenology of additional VLLs is expected to be similar to excited leptons or can differ if 
the model predicts additional particles in the spectrum. The VLLs can also modify the SM Higgs boson decay to di-photon mode.
The VLLs are however less constrained than chiral fourth generation of SM fermions, typically from electroweak precision observables and Higgs signal strengths \cite{Bahrami:2015mwa}. The VLLs have also been studied in the context of 
dark matter (DM) phenomenology and  several DM and collider searches have been performed in the framework containing SM augmented with Higgs singlet \cite{Bell:2019mbn}, Higgs doublet \cite{Osoba:2012pc,Garg:2013rba,Angelescu:2016mhl,Angelescu:2015uiz}, Higgs triplet \cite{Bahrami:2013bsa,Bahrami:2015mwa,Bahrami:2015rqa}, left right symmetric model
\cite{Chakdar:2013tca,Bahrami:2016has,Patra:2017gak} extended by one or more generations of VLLs. 

In the present analysis, we study an $S_3$-symmetric two Higgs doublet model (2HDM) \cite{Das:2017zrm,Cogollo:2016dsd}, along with two generations of VLLs. The addition of two generations of VLLs in this model helps guarantee an $S_3$-symmetric Yukawa Lagrangian providing an aesthetic picture to their inclusion. The main motivation of $S_3$-symmetric 2HDM is to 
provide proper mass hierarchy and mixing among the SM fermions. Besides, the $S_3$-symmetric 2HDM 
incorporates a 125 GeV SM-like Higgs boson in a very simple and natural way, unlike in a general 2HDM \cite{Das:2017zrm,Cogollo:2016dsd}. In our model we impose an additional $Z_2$ symmetry, under which all the SM fermions are even and the VLLs are odd  \cite{Chakraborty:2021tdo}. Thus the mixing between the SM fermions and VLLs is forbidden throughout the analysis, and the lightest neutral VLL serves as a viable DM candidate which satisfies correct relic density, direct detection cross-sections and thermally averaged annihilation cross-sections in indirect detection obtained from the experiments \cite{Chakraborty:2021tdo}.  
A rigorous collider analysis of the multi-leptons + missing transverse energy final state at high-luminosity (HL) LHC can be found in one of our recent studies \cite{Chakraborty:2021tdo} where we found that the LHC provided us with a limited sensitivity for 
the VLLs for large masses as well as when the spectrum satisfying DM results demanded a compressed spectrum. Such 
spectrums will be more likely to be observable in a cleaner environment of an electron-positron collider such as the 
International Linear Collider (ILC) \cite{Behnke:2013xla,Baer:2013cma}. The ILC will be an invaluable machine with several exciting physics studies
and is hence proposed to run at several center of mass energies, each driven by the physics study it aims to achieve. 
For our analysis, we have chosen the high energy option of $\sqrt{s}=1$ TeV 
that allows a larger phase space to produce heavier VLLs. In this work, we therefore study leptonic and hadronic states
with missing transverse energy at 1 TeV ILC to highlight the sensitivity. We observed that the benchmarks with high DM 
masses or with a compressed particle spectrum which were challenging to probe owing to small signal cross-sections at 
HL-LHC are easily discernible with high significance in specific final states involving hadronic final states. 
To perform our collider analysis we select some benchmark points with low, medium and high DM masses from the 
multi-dimensional parameter space satisfying the theoretical, experimental and DM constraints. There exist several searches 
by ATLAS and CMS involving di-lepton \cite{Aad:2019vnb}, four leptons \cite{Aaboud:2018zeb,Aad:2021lzu} and multi jets \cite{ATLAS-CONF-2019-040} along with missing transverse energy in the final states. We have validated all the benchmark 
points with the limits arising out of these existing studies. To optimize the signal over the SM backgrounds, for each channel 
we have performed a cut-based analysis and also shown the possible improvement in the analysis employing 
machine learning with a sophisticated multivariate technique.

The paper is organized as follows. In section \ref{model}, we discuss the relevant scalar and Yukawa sector of the model. 
In section \ref{sec:collider}, we present a collider analysis of the leptonic and hadronic final states along with missing transverse energy. Finally we summarize and conclude in Section \ref{conc}.


\section{Model}
\label{model}
We work in the $S_3$-symmetric 2HDM which contains two generations of VLLs. Including two generations of VLLs instead of one allows one to write a Yukawa Lagrangian fully $S_3$-symmetric. Each generation of VLL comprises of one left-handed lepton doublet $L_{L_i}'$, one right-handed charged lepton singlet $e_{R_i}'$ and one right-handed singlet neutrino $\nu_{R_i}'$, accompanied by their mirror counter parts with opposite chirality, {\em i.e.} $L_{R_i}'', e_{L_i}''$ and $\nu_{L_i}''$ with $i=1,2.$ The quantum numbers for the SM and beyond Standard Model (BSM) particles are shown in Table \ref{quantum}, while Table \ref{quantum-S3} shows the $S_3$ quantum numbers of the particles.   Two Higgs doublets $\phi_1$ and $\phi_2$ together form an $S_3$-doublet. In Table \ref{quantum}, $Q_{iL}, L_{iL}$ are the SM left-handed quark and lepton doublets, while $u_{iR},~ d_{iR},~ e_{iR}$ are the right-handed up-type, down-type quark and charged lepton singlets respectively for $i=1,2,3$.
\begin{table}[htpb!]
\begin{center}
\begin{tabular}{|c|c|c|c|c|}
\hline
\hspace{7mm} Fields \hspace{7mm} &  \hspace{5mm} $SU(2)_L$ \hspace{5mm} & \hspace{5mm} $SU(3)_C$ \hspace{5mm} & \hspace{5mm} $U(1)_Y$~~ \hspace{5mm} & \hspace{5mm} $Z_2$ ~~\hspace{5mm}\\
\hline
\hline
$\phi_1$ & 2& 1& +1 & 1\\
\hline
$\phi_2$ &2 &1  & +1 & 1\\
\hline
$Q_{iL}, ~~ i =1,2,3$ &2 &3  & $+\frac{1}{3}$ &1\\
\hline
 $u_{iR}, ~~ i =1,2,3$ &1 & 3 & $+\frac{4}{3}$ & 1\\
 \hline 
 $d_{iR}, ~~ i =1,2,3$ &1 & 3 & $-\frac{2}{3}$ & 1\\
 \hline
$L_{iL}, ~~ i =1,2,3$ & 2& 1&  $-1$ & 1 \\
\hline
$e_{iR},  ~~ i =1,2,3$ & 1& 1 & $-2$ &  1\\ 
\hline
$L_{L_i}',~~ i =1,2$ &2 & 1& -1 & -1\\
\hline
$L_{R_i}'',~~ i =1,2$ &2 & 1& -1 & -1 \\
\hline
$e_{R_i}', ~~ i =1,2$ &1 & 1& -2 & -1\\
\hline
$e_{L_i}'',~~ i =1,2$ &1 & 1& -2 & -1\\
\hline
$\nu_{R_i}',~~ i =1,2$ &1 & 1& 0 & -1\\
\hline
$\nu_{L_i}'', ~~ i =1,2$ &1 & 1& 0 & -1\\
\hline
\end{tabular}
\end{center}
\caption{$SU(2)_L \times SU(3)_C \times U(1)_Y \times Z_2$ quantum numbers assigned to the particles in the model.}
\label{quantum}
\end{table}
\begin{table}[htpb!]
\begin{center}
\begin{tabular}{|c||c|}
\hline
Particles  &   $S_3$  \\ \hline \hline
$\begin{pmatrix}
\phi_1 \\
\phi_2
\end{pmatrix}$;  \hspace{2mm} 
$\begin{pmatrix}
Q_{1L} \\
Q_{2L}
\end{pmatrix}$;  \hspace{2mm} 
$\begin{pmatrix}
u_{1R} \\
u_{2R}
\end{pmatrix}$;  \hspace{2mm}
$\begin{pmatrix}
d_{1R} \\
d_{2R}
\end{pmatrix}$;  \hspace{2mm}
$\begin{pmatrix}
\ell_{1L} \\
\ell_{2L}
\end{pmatrix}$; \hspace{2mm}
$\begin{pmatrix}
e_{1R} \\
e_{2R}
\end{pmatrix}$
& 2  \\ \hline
$Q_{3L}$, $u_{3R}$, $d_{3R}$, $\ell_{3L}$, $e_{3R}$ & 1 \\ \hline \hline 
$\begin{pmatrix}
L_{L_1}' \\
L_{L_2}' \\
\end{pmatrix}$;  \hspace{5mm}
$\begin{pmatrix}
L_{R_1}'' \\
L_{R_2}'' \\
\end{pmatrix}$;  \hspace{5mm}
$\begin{pmatrix}
e_{R_1}' \\
e_{R_2}' \\
\end{pmatrix}$;  \hspace{5mm}
$\begin{pmatrix}
e_{L_1}'' \\
e_{L_2}'' \\
\end{pmatrix}$; \hspace{5mm}
$\begin{pmatrix}
\nu_{R_1}' \\
\nu_{R_2}' \\
\end{pmatrix}$;   \hspace{5mm}
$\begin{pmatrix}
\nu_{L_1}'' \\
\nu_{L_2}'' \\
\end{pmatrix}$  \hspace{5mm} & 2 \\
\hline 
\hline
\end{tabular}
\end{center}
\caption{$S_3$ quantum number assigned to the particles in the model.}
\label{quantum-S3}
\end{table}
 
\label{Model}
\subsection{Scalar and Yukawa Lagrangian}
 Two $SU(2)_L$ doublets $\phi_1$ and $\phi_2$ in $S_3$-symmetric 2HDM with hypercharge $Y= + 1$ , jointly behave like a doublet under $S_3$-symmetry, {\em i.e.}
$ \Phi = \begin{pmatrix}
\phi_1 \\
\phi_2
\end{pmatrix} $. 

The neutral components of $\phi_i$ acquire vacuum expectation value (responsible for the spontaneous symmetry breaking (SSB) of SM gauge symmetry). The doublets can be written as shown below,
\begin{eqnarray}
\phi_i = \begin{pmatrix}
\phi_i^+ \\
\frac{1}{\sqrt{2}} (v_i + h_i + i \rho_i)
\end{pmatrix}
\end{eqnarray}
Here $v_i$'s are VEVs of two doublets with $v_1 = v \cos \beta , ~v_2 = v \sin \beta$ and $v = \sqrt{v_1^2 + v_2^2} = 246$ GeV. The ratio of two vacuum expectation values can be denoted by $\tan \beta$ , {\em i.e.} $\tan \beta = \frac{v_2}{v_1}$.

The most general renormalisable scalar potential for $S_3$-symmetric 2HDM can be written as the sum of $V_2(\phi_1, \phi_2)$ and $V_4(\phi_1, \phi_2)$  \cite{Das:2017zrm} :
\bea
V(\phi_1, \phi_2) = V_2(\phi_1, \phi_2) + V_4(\phi_1, \phi_2)
\label{V-tot}
\eea
with
\begin{eqnarray}
 V_2(\phi_1, \phi_2)&=& m_{11}^2 (\phi_1^\dagger\phi_1) + m_{22}^2 (\phi_2^\dagger\phi_2) - \{m_{12}^2 (\phi_1^\dagger\phi_2) + {\rm h.c.}\} 
\label{V2}
\end{eqnarray}
and
\begin{eqnarray}
 V_4(\phi_1, \phi_2)&=&  \lambda_1 (\phi_1^\dagger\phi_1+\phi_2^\dagger\phi_2)^2  +\lambda_2 (\phi_1^\dagger\phi_2 -\phi_2^\dagger\phi_1)^2 \nonumber \\
 && + \lambda_3
\left\{(\phi_1^\dagger\phi_2+\phi_2^\dagger\phi_1)^2
+(\phi_1^\dagger\phi_1-\phi_2^\dagger\phi_2)^2\right\} \,.
\label{V4}
\end{eqnarray}

In Eq.(\ref{V2}) and Eq.(\ref{V4}), the subscripts denote the dimensionality of the terms. The hermiticity of the scalar potential in Eq.(\ref{V4}), forces the quartic couplings $\lambda_1, \lambda_2$ and $\lambda_3$ to be real. In $V_2(\phi_1, \phi_2)$, $m_{11}^2, m_{22}^2$ are real, $m_{12}^2$ can be complex in principle. In this analysis, we shall not consider $m_{12}^2$ to be complex to circumvent $CP$-violation. The configuration $m_{11}^2 = m_{22}^2$ along with $m_{12}^2 =$ 0 makes the quadratic part of the potential $S_3$-symmetric. At the same time this condition results in a massless heavy Higgs boson \cite{Das:2017zrm}. Thus to avoid any other massless heavy Higgs boson apart from the Goldstone bosons, we adhere to : $m_{11}^2 = m_{22}^2$ and $m_{12}^2 \neq$ 0.  Thus the value of $\tan \beta$ is fixed to 1, following the minimisation conditions of the  scalar potential in Eq.(\ref{V-tot}) \cite{Das:2017zrm}.

The particle spectrum of this model comprises of SM-like Higgs ($h$), heavy CP-even Higgs ($H$), pseudoscalar Higgs ($A$) and charged Higgs ($H^{\pm}$). The {\em alignment limit}, in which $h$ resembles SM Higgs boson, is naturally achieved in this model \cite{Das:2017zrm}.



The most general Yukawa Lagrangian involving two generations of VLLs is given by,
\begin{eqnarray}
\mathcal{L}_{\rm{Yuk}} &=& -M_{1}\bar{L}_{L_1}^{'}L_{R_1}^{''}
-M_{2}\bar{L}_{L_1}^{'}L_{R_2}^{''}-M_{3}\bar{L}_{L_2}^{'}L_{R_1}^{''}-M_{4}\bar{L}_{L_2}^{'}L_{R_2}^{''} \nonumber \\
&& -\frac{1}{2}M_{5}\bar{\nu^c}_{L_{1}}^{''}\nu_{L_{1}}^{''}-\frac{1}{2}M_{6}\bar{\nu^c}_{L_{2}}^{''}\nu_{L_{2}}^{''}
-\frac{1}{2}M_{7}\bar{\nu^c}_{R_{1}}^{'}\nu_{R_{1}}^{'}-\frac{1}{2}M_{8}\bar{\nu^c}_{R_{2}}^{'}\nu_{R_{2}}^{'} -M_{9}\bar{\nu}_{L_{1}}^{''}\nu_{R_{1}}^{'} \nonumber \\
&& -M_{10}\bar{\nu}_{L_{1}}^{''}\nu_{R_{2}}^{'}- M_{11}\bar{\nu}_{L_{2}}^{''}\nu_{R_{1}}^{'}-M_{12}\bar{\nu}_{L_{2}}^{''}\nu_{R_{2}}^{'}  -M_{L_{1}}\bar{e}_{L_{1}}^{''}e_{R_{1}}^{'}-M_{L_{2}}\bar{e}_{L_{2}}^{''}e_{R_{2}}^{'} \nonumber \\
&&-M_{L_{3}}\bar{e}_{L_{1}}^{''}e_{R_{2}}^{'}-M_{L_{4}}\bar{e}_{L_{2}}^{''}e_{R_{1}}^{'} \nonumber \\
&&-y_2[(\bar{L}_{L_1}^{'}\tilde{\phi_2}+\bar{L}_{L_2}^{'}\tilde{\phi_1})\nu_{R_{1}}^{'} +  
(\bar{L}_{L_1}^{'}\tilde{\phi_1}-\bar{L}_{L_2}^{'}\tilde{\phi_2})\nu_{R_{2}}^{'}]
-y_4[(\bar{L}_{R_1}^{''}\tilde{\phi_2}+\bar{L}_{R_2}^{''}\tilde{\phi_1})\nu_{L_{1}}^{''}\nonumber \\
&& + (\bar{L}_{R_1}^{''}\tilde{\phi_1}-\bar{L}_{R_2}^{''}\tilde{\phi_2})\nu_{L_{2}}^{''}]
-y_2^{'}[(\bar{L}_{L_1}^{'}\phi_2+\bar{L}_{L_2}^{'}\phi_1)e_{R_{1}}^{'}+
(\bar{L}_{L_1}^{'}\phi_1-\bar{L}_{L_2}^{'}\phi_2)e_{R_{2}}^{'}] \nonumber \\
&& -y_4^{'}[(\bar{L}_{R_1}^{''}\phi_2+\bar{L}_{R_2}^{''}\phi_1)e_{L_{1}}^{''}+
(\bar{L}_{R_1}^{''}\phi_1-\bar{L}_{R_2}^{''}\phi_2)e_{L_{2}}^{''}]
+ {\rm h.c.} 
\label{Yukawa-pot}
\end{eqnarray}

Here the charge conjugated fields are denoted with superscript "c" in Eq.(\ref{Yukawa-pot}). In presence of exact $S_3$-symmetry, the masses of the VLLs will be proportional to the product of Yukawa coupling and the electroweak vacuum expectation value (VEV), which in turn will lead to non-perturbative Yukawa couplings (for vector lepton masses $\sim$ 1 TeV). Thus we introduce Dirac and Majorana mass terms in  Eq.(\ref{Yukawa-pot}), that break $S_3$-symmetry softly, while the rest of the terms in  Eq.(\ref{Yukawa-pot}) are $S_3$-symmetric. 

Two generations of VLLs comprise of eight neutral and four charged  flavor eigenstates. Thus we can construct eight neutral mass eigenstates ($N_i , i = 1$-8) and four charged mass eigenstates ($E_i^+ , i = 1$-4) out of the aforementioned flavor eigenstates. The unbroken $Z_2$ symmetry in the model allows the lightest neutral state of the VLLs to act as the DM candidate. It also ensures that the mixing of VLL with SM fermions is also prohibited. We do note that the model can incorporate tiny 
neutrino masses radiatively through contributions from the $Z_2$ odd fermions. We however focus on the collider 
signals of the VLLs at the ILC and leave that study for future considerations. The mass matrices for the neutral and charged fermions and the details of their diagonalisation can be found in \cite{Chakraborty:2021tdo}. 

\section{Collider Searches}
\label{sec:collider}
By the virtue of $Z_2$-symmetry, the lightest neutral VLL $N_1$ cannot decay and becomes a possible DM candidate. 
To explore the model parameter space compatible with relic density ($\Omega_{\rm DM} h^2$)\footnote{$\Omega_{\rm DM}$ 
is defined as the ratio of non-baryonic DM density to the critical density of the universe and $h$ is the reduced Hubble 
parameter (not to be confused with SM Higgs $h$).}, direct and indirect DM searches, we implement the model Lagrangian
in \texttt{FeynRules}~\cite{Alloul:2013bka} to generate the interaction vertices and mass matrices. The \texttt{CALCHEP}~\cite{Belyaev:2012qa} compatible model files obtained from \texttt{FeynRules} is then included in \texttt{micrOMEGAs}~\cite{Belanger:2014vza}, which helps us to calculate the DM observables like relic density ($\Omega_{\rm DM} h^2$),
spin-dependent ($\sigma_{\rm SD}$) and spin-independent  ($\sigma_{\rm SI}$) cross-sections, thermally averaged 
annihilation cross-sections ($ \langle \sigma v \rangle$), etc. We choose four  representative benchmark points BP1 BP2, 
BP3 and BP4 according to low, medium and high DM masses, that we found consistent with observed relic abundance 
obtained from the PLANCK experiment \cite{Aghanim:2018eyx}, and also allowed by the stringent bounds coming from the 
direct detection experiments like LUX \cite{Akerib:2016vxi} for spin-independent cross-sections and
PICO \cite{Amole:2019fdf} for spin-dependent cross-sections,  and from the indirect detection bounds coming from 
FERMI-LAT \cite{Daylan:2014rsa}, MAGIC \cite{Ahnen:2016qkx} and PLANCK \cite{Aghanim:2018eyx} experiments. 
The masses of the neutral ($N_i$s) and charged ($E_i^+$s) VLLs for our chosen benchmark points can be found in 
Table \ref{table:masses}. All these four benchmark points represent model parameters which satisfy theoretical constraints 
like stability of the scalar potential, perturbativity and constraints coming from electroweak precision data and Higgs signal 
strengths\footnote{More details can be found in our earlier work~\cite{Chakraborty:2021tdo}.}. $\Omega_{\rm DM} h^2,~\sigma_{\rm SD},~\sigma_{\rm SI},~\langle \sigma v \rangle$ and dominant annihilation modes for indirect detection {\footnote{The 6th and 7th columns of Table \ref{table:dmbp}  actually refer to the $<\sigma v>$ relevant for indirect detection, and, the corresponding annihilation channels respectively. The aforementioned indirect detection annihilation cross sections alone cannot lead to an estimate for the relic density since there might be other annihilation/coannihilation channels entering the relic density calculation.}} along with corresponding dark matter  masses for the aforementioned benchmark points (BPs) are tabulated in Table \ref{table:dmbp}. Our choice of BP's 
is envisaged to cover varied and complementary features of the model. For example, BP1 corresponds to low DM mass, BP2 corresponds to a slightly heavier DM mass with substantial mass splitting with the charged VLLs, BP3 corresponds to a 
{\it compressed~spectrum} where the mass difference between the components of the VLL's are $\sim$ 20 GeV while BP4 corresponds to an overall heavy spectrum with higher DM mass which renders them  close to threshold value of the ILC 
center of mass energy.

\begin{table}[htbp!]
	\centering
	\resizebox{16cm}{!}{
	\begin{tabular}{|p{2.1cm}|c|c|c|c|c|c|c|c|c|c|c|c|c|c|}
		\hline
		Benchmark  & $M_{N_1}$ & $M_{N_2}$  & $M_{N_3}$ & $M_{N_4}$  & $M_{N_5}$ & $M_{N_6}$ & $M_{N_7}$ &  $M_{N_8}$
		&$M_{E^{\pm}_1}$ & $M_{E^{\pm}_2}$ & $M_{E^{\pm}_3}$ & $M_{E^{\pm}_4}$\\ 
	Points &$(\rm GeV)$ &$(\rm GeV)$& $(\rm GeV)$ & $(\rm GeV)$ &  $(\rm GeV)$ & $(\rm GeV)$ & $(\rm GeV)$ & $(\rm GeV)$ & $(\rm GeV)$ &
		 $(\rm GeV)$ & $(\rm GeV)$ & $(\rm GeV)$\\ \hline
		BP1  & 81.3  & 86.9 & 119.3 & 154.4 &  211.6 &   268.7 & 688.4 & 856.9 & 171.0 & 211.8 & 260.0 & 322.0 \\ \hline
		BP2 & 193.8 & 204.8 & 239.8 & 245.7 & 268.3 &  274.9  & 454.5 & 494.7 & 280.2 & 313.0 & 356.8 & 398.5  \\ \hline
		BP3 & 261.4 & 262.5 & 263.2 & 263.4 & 264.0 & 297.1 & 444.9 & 505.5 & 280.2 & 313.0 & 356.7 & 398.5 \\ \hline
		BP4 & 402.8 & 456.9 & 461.3 & 466.1 & 467.3 &
		508.0 & 518.3 & 653.7 & 486.2 & 539.4 & 592.3 & 657.1 \\ \hline
		
	\end{tabular}}
	\caption{Masses of neutral and charged VLLs for four benchmarks.}
	\label{table:masses}
\end{table}

\begin{table}[htbp!]
\centering
	\resizebox{16cm}{!}{
		\begin{tabular}{|p{2cm}|c|c|c|c|p{3cm}|p{3cm}|}
		\hline
		Benchmark Points &  $M_{\rm DM}$  & $\Omega_{\rm DM} h^2$ & $\sigma_{\rm SD}$  & $\sigma_{\rm SI}$ &
		Annihilation cross-section  & Annihilation mode \\ 
		& & & & & (for indirect detection only) & (for indirect detection only) \\
		 &   $(\rm GeV)$ &  & ($cm^2$) &  ($cm^2$) & $\langle \sigma v \rangle$ ($cm^3/s$) &  \\ \hline
 BP1 & 81.3    & $1.04\times 10^{-1}$ & $3.4\times10^{-42} $ & $4.4\times10^{-49}$ &$2.41\times10^{-28}$ & $W^{+}W^{-}$ (100\%) \\ \hline
 
 BP2 & 193.8    & $9.36\times 10^{-4}$ & $5.98\times10^{-42}$ & $2.54\times10^{-49}$ &$3.79\times10^{-26}$ & $ZZ$ ( 53.8\%)\\
 & & & & & &  $W^{+}W^{-}$ (46.1\%) \\ \hline
 
 BP3 & 261.4 & $4.14\times 10^{-3}$ & $2.57\times10^{-41}$& $1.06\times10^{-46}$& $2.49\times10^{-26}$ &  $ZZ$ (55.5\%) \\
 & & & & & & $W^{+}W^{-}$ (44.3\%) \\ \hline
 
 BP4 &  402.8 & $2.97\times 10^{-4}$ & $7.16\times10^{-41}$ & 
 $9.86\times 10^{-49}$ & $1.05\times 10^{-26}$ & $ZZ$ (52.0\%) \\
 & & & & & & $W^{+}W^{-}$ (46.3\%) \\ \hline
 
			\end{tabular}}
			
	\caption{ DM masses along with
		DM relic density, spin-dependent, spin-independent cross-sections, thermally averaged indirect detection annihilation cross-sections and corresponding dominant annihilation modes  for four benchmarks.}
	\label{table:dmbp}
\end{table}


We focus on the $\sqrt{s} = 1$ TeV of ILC and present our analysis for some specific processes in the $S_3$-symmetric model which gives us  the semi-leptonic $ 1 \ell + 2j + \mET$, fully leptonic $2 \ell + \mET$ and  $4 \ell + \mET$\footnote{Here $\ell$ represents electron or muon and $\mET$ denotes missing transverse energy} and the fully hadronic $ 4j + \mET$ final states. The $ 1 \ell + 2j + \mET$ and $ 4j + \mET$  channel containing multi-jets prove to be promising signals at the ILC, compared to LHC where huge SM backgrounds would supersede the signal. The spectrum with higher DM mass also proved hard to search at LHC \cite{Chakraborty:2021tdo} even with 
high integrated luminosity, since corresponding signal cross-section was too small to yield significant signal significance. 

To generate the signal and SM background at leading order (LO), we use the public package {\tt MG5aMC@NLO}~\cite{Alwall:2014hca}. We first use the following $\it{acceptance~cuts}$ : 
\begin{eqnarray}
p_{T}^j > 20 ~{\rm GeV},~~~~|\eta_j| < 5.0, \nonumber \\
p_{T}^{\ell} > 10 ~{\rm GeV},~~~~|\eta_\ell| < 2.5, \nonumber \\
\Delta \, R_{ij}>0.4, ~ {\rm with}~~ i,j = \ell, ~{\rm jets}.
\label{basic_cuts}
\end{eqnarray}

Here $p_T^{j(\ell)}, ~|\eta_{j(\ell)}|$ are the transverse momentum and pseudo-rapidity of jets (leptons). $\Delta R_{ij}$ is defined as : $\Delta R_{ij} = \sqrt{\Delta \eta_{ij}^2 + \Delta \phi_{ij}^2}$, where $\Delta \eta_{ij}$ ($\Delta \phi_{ij}$) is the difference between the pseudo-rapidity (azimuthal angles) of $i^{\rm th}$ and $j^{\rm th}$ particle in the final state.
Subsequent decays of the unstable particles are incorporated in {\tt Pythia8}~\cite{Sjostrand:2014zea}. To emulate the detector effects into the analysis, we pass the signal and background events  in  {\tt Delphes-3.4.1} ~\cite{deFavereau:2013fsa} using the default ILD detector simulation card. We note that the results obtained from traditional cut-based analysis are improved further by using Decorrelated Boosted Decision Tree (BDTD) algorithm embedded in TMVA (Toolkit for Multivariate Data Analysis) \cite{Hocker:2007ht} platform.  The signal significance $\mathcal{S}$ can be calculated in both cut-based and multivariate analysis using $\mathcal{S} = \sqrt{2\Big[(S + B) \log\Big(\frac{S + B}{B}\Big)- S\Big]}$, with $S (B)$ denoting the number of signal (background) events surviving the cuts on relevant kinematic variables.

\subsection{$2\ell+\mET$ final state}
\label{2lmet}
We begin with the leptonic signal consisting of two charged leptons. The final state contains same or different flavour 
and opposite sign (OS) di-leptons along with $\mET$.  For the benchmark points BP1, BP2 and BP4 dominant contribution 
to the $2\ell+\mET$ final state comes from $e^+ e^- \to E_1^+ E_1^- \to W^{+} W^{-} \mET$ channel, where 
$W^{\pm}$ is assumed to decay leptonically. But for BP3, the contribution comes from three body decay, namely 
$e^+ e^- \to E_1^+ E_1^- \to \ell^{+} \ell^{-} \mET$. This is due to the fact that mass difference between 
$E_1^{\pm}$ and $N_1$ is much less than the mass of $W^{\pm}$. Here we consider the following processes 
that can lead to the $2\ell+\mET$ final state:
\begin{eqnarray} 
&& e^+ e^- \to E_i^+ E_j^-, ~ E_{i,j}^\pm \rightarrow \ell^\pm N_1, \nonumber \\
&& e^+ e^- \to N_k N_1 , ~ N_k \to \ell^+ \ell^- N_1
 \label{inclusive}
\end{eqnarray}
where $i,j = 1...4$ and $ k,m = 1...8$. These processes can give rise to the $2\ell+\mET$ final state depending on the mass spectrum of the VLL's in individual benchmarks. Here we choose the same or different flavour and opposite sign (OS) di-leptons in such a way that the leading and sub-leading leptons have the transverse momenta greater than 10 GeV, {\em i.e.} $p_T^{\ell_1, \ell_2} >10$ GeV. At the same time, we reject any third lepton in the final state. Since our signal does not contain any jet, we veto the light jets as well as $b$ jets.

The dominant background comes from the $ ~e^+ e^- \to  \ell^{+}  \ell^{-} + \mET$ final state comprising of the following possible subprocesses that lead to a similar final state :
\begin{itemize}
\item $e^+ e^- \to W^+ W^-$; $W^+ \to \ell^+ ~\nu_\ell, ~ W^-  \to \ell^- ~\bar{\nu}_\ell$,
\item $e^+ e^- \to ZZ$; $Z \to \ell^+ \ell^-, ~ Z  \to \nu_\ell ~\bar{\nu}_\ell$,
\item $e^+ e^- \to W^+ W^- Z$; $W^+ \to \ell^+ ~\nu_\ell, ~ W^-  \to \ell^- ~\bar{\nu}_\ell, ~ Z  \to \nu_\ell ~\bar{\nu}_\ell$,
\item $e^+ e^- \to ZZZ$; $Z \to \ell^+ \ell^-, ~ Z  \to \nu_\ell ~\bar{\nu}_\ell, ~ Z  \to \nu_\ell ~\bar{\nu}_\ell$.
\end{itemize}
 
 A promising feature at the ILC would be the use of polarized beams which can affect specific physics studies in which 
 certain chirality in currents are favored/conserved. To see our signal in the same spirit we calculate the signal and 
 background cross sections corresponding to both polarised ([80$\%$ left polarised $e^-$, 30$\%$ right polarised $e^+ $ beam] and [80$\%$ right polarised $e^-$ and unpolarised $e^+$ beam]) and unpolarised $e^+$-$e^-$ beams \cite{Beyer:2021nbp} which are tabulated 
 in Table \ref{tab:bp_ilc}. 
 From now on, we shall only present the distributions and calculate the signal significances for the unpolarised incoming beams using cut-based as well as BDT analysis.

\begin{table}[!hptb]
\centering
\resizebox{13cm}{!}{
 \begin{tabular}{|c|c|c|c|}
  \hline \hline
  & Cross section for  &  Cross section for   & Cross section for \\
  & ($P_{e^-},P_{e^+} = 0,0)$ & ($P_{e^-},P_{e^+} = 80\%L,30\%R)$ &  ($P_{e^-},P_{e^+} = 80\%R,0)$  \\
  & (in fb) &  (in fb) & (in fb) \\
  \hline
  \multicolumn{4}{|c|}{Signal benchmarks}  \\ \hline \hline
  BP1 & 6.8 & 11.77 & 5.11 \\ \hline
  BP2 & 5.28 & 9.49 & 3.60 .
  \\ \hline
  BP3  & 4.42 & 7.95 & 3.01 \\ \hline
  BP4 & 0.71 & 1.29 & 0.47 \\ \hline
 
   \multicolumn{4}{|c|}{Background }  \\ \hline \hline
  $e^+ e^- \rightarrow 2\ell+\mET$ & 420.22 & 922.3 & 88.81 \\ \hline
  \hline
 \end{tabular}}
	\caption{ The leading order (LO) effective cross-sections of
	the signal and background for $2\ell+ \mET$ final state at 1 TeV ILC using unpolarised and polarised incoming beams. $P_{e^-},P_{e^+}$ denotes the polarisation for the $e^-$ and $e^+$ beam respectively. L and R is used to denote whether the beam is left or right polarised.}
	\label{tab:bp_ilc}
 \end{table}

We carry out the cut-based analysis by looking at some relevant kinematic variables which can help design proper cuts ($A_1, A_2, A_3$) on them to improve the signal over the backgrounds.

\begin{figure}[htpb!]{\centering
\subfigure[]{
\includegraphics[height = 5.5 cm, width = 8 cm]{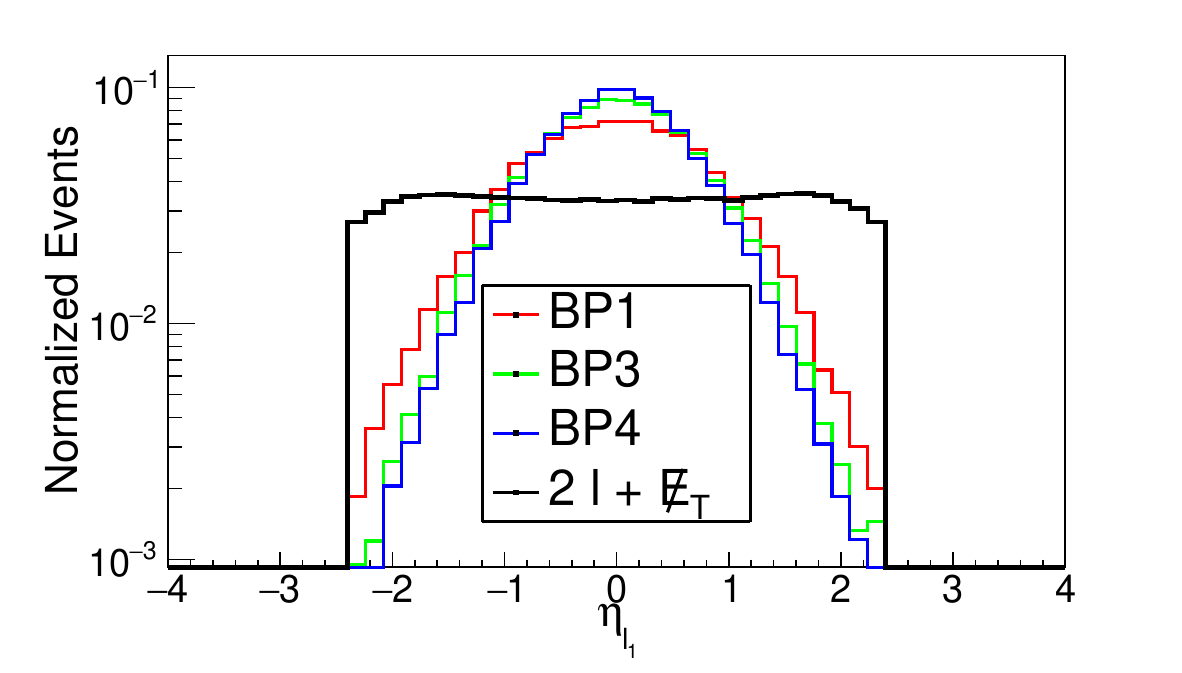}} 
\subfigure[]{
\includegraphics[height = 5.5 cm, width = 8 cm]{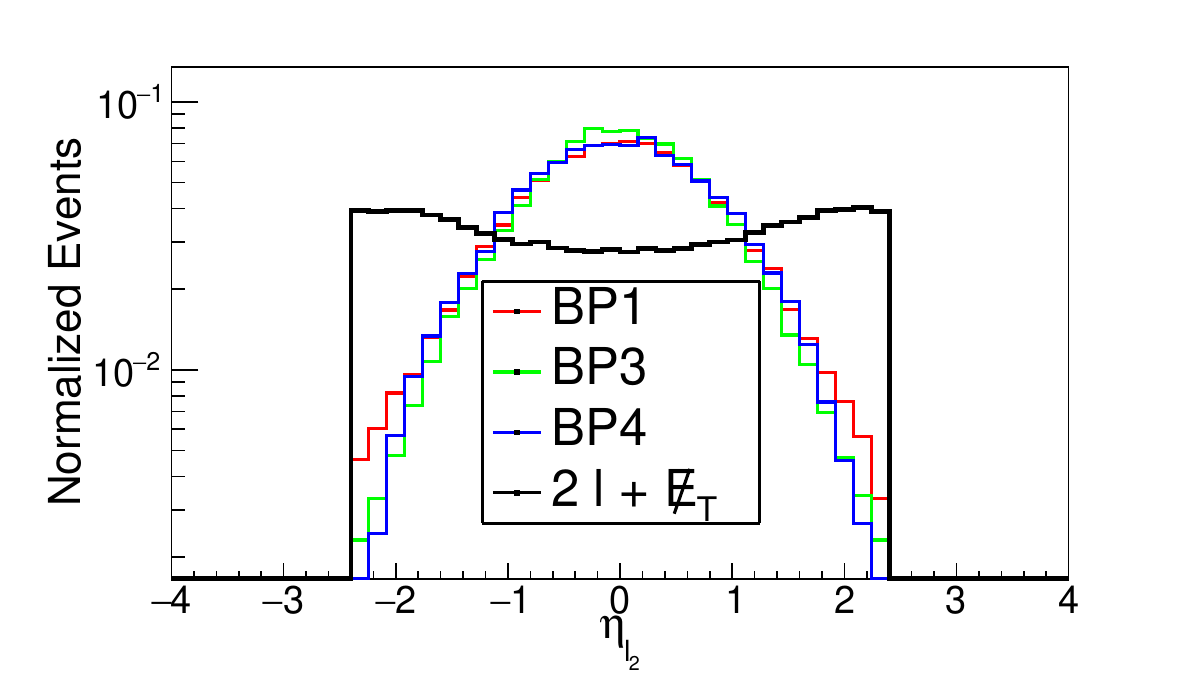}} \\
\subfigure[]{
\includegraphics[height = 5.5 cm, width = 8 cm]{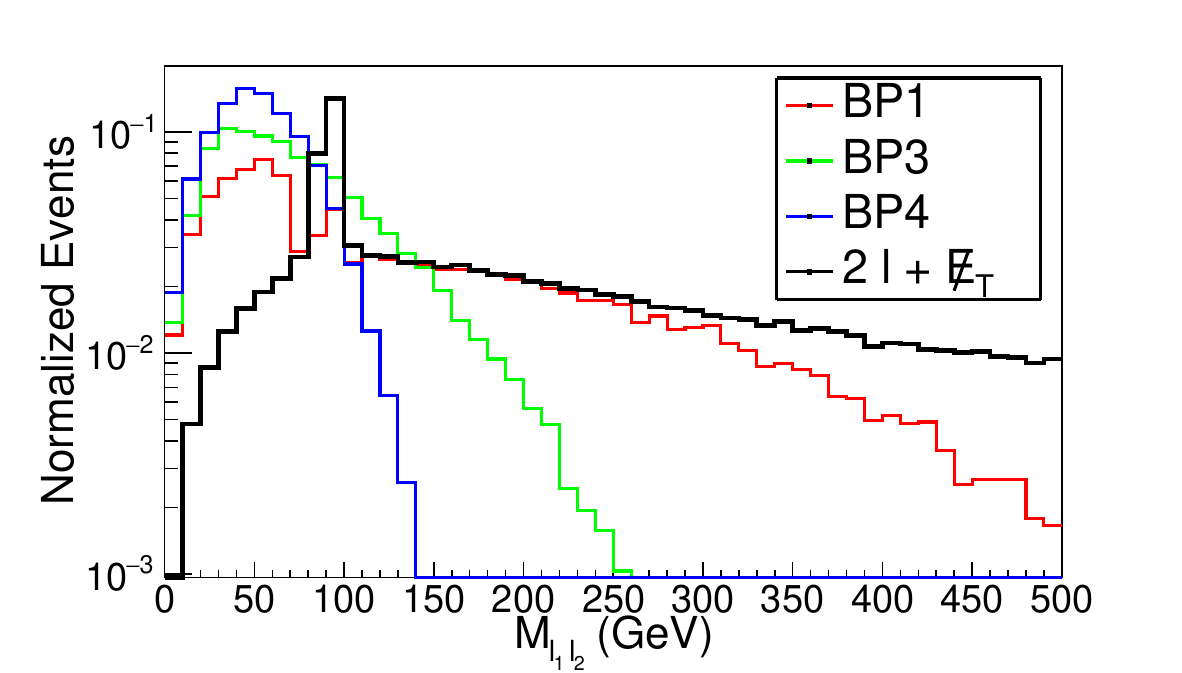}}
\subfigure[]{
\includegraphics[height = 5.5 cm, width = 8 cm]{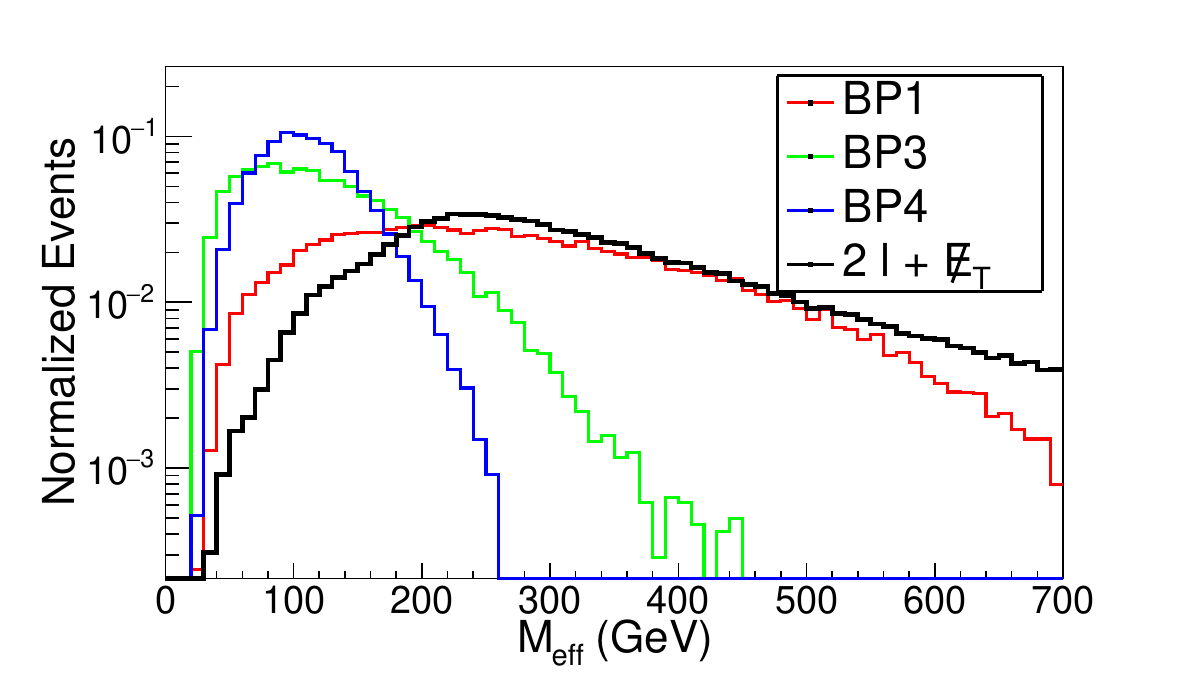}}
} \\
\caption{ Normalized distributions of $\eta_{\ell_1},~\eta_{\ell_2}, ~M_{\ell_1 \ell_2}, ~M_{\rm eff}$ for $2 \ell + \mET$ channel at 1 TeV ILC. }
\label{distribution-dilep}
\end{figure}

\begin{itemize}
 
 \item $A_1$ : We depict the normalized pseudorapidity distributions of the leading and sub-leading leptons in Fig.~\ref{distribution-dilep}(a) and Fig.~\ref{distribution-dilep}(b). For the background, the leptons can be produced via s-channel exchange of $\gamma, Z$ as well as t-channel exchange of $\nu$'s, which results in the peaks at higher $\eta$ values. However, for the signal, the leptons are produced from the decay of the $W^{\pm}$, which are generated from the decay of heavier VLLs, produced via s-channel exchange of $\gamma, Z$. As a result, the $\eta$ distribution for the signal is more centrally peaked. We note that, choosing $|\eta_{l_{1,2}}| < 1.0$ helps to reduce the background significantly. 
 
\item $A_2$ : The normalized distribution of the invariant mass of the opposite sign (OS) lepton pair (same or different flavor) $M_{\ell^+ \ell^-}$ is shown in Fig.~\ref{distribution-dilep}(c). Since the $2\ell+\mET$ background consists of the contributions from $ZZ$ and $\gamma^{*}Z$, to exclude the Z-peak, we reject events which lie within the window : $ |M_{\ell^+ \ell^-} - M_Z| < 15$ GeV. At the same time, we also demand $M_{\ell^+ \ell^-} > 12$ GeV
to reduce the $\gamma^{*}Z$ background contribution.

\item $A_3$ : The variable $M_{\rm eff}$ is constructed as the scalar sum of the lepton $p_T$ and $\mET$. The distribution is shown in Fig.~\ref{distribution-dilep}(d). Instead of giving cuts on the lepton $p_T$ and $\mET$ separately, it is useful to put a cut on $M_{\rm eff}$ which helps to reject the background more efficiently. Since for BP3 and BP4, the mass difference between the charged and neutral component of the VLL's are smaller compared to BP1 and BP2, the lepton $p_T$ is less. As a result, $M_{\rm eff}$ peaks at a smaller value for BP3 and BP4. We impose an upper cut : $M_{\rm eff} < $ 500 (350) GeV for BP1 and BP2 (BP3 and BP4) to reduce the background. 
\end{itemize}

We tabulate the number of signal and background events surviving after the application of each cut for each benchmark at an integrated luminosity 100 fb$^{-1}$ in Table~\ref{tab:sig_ilc_2lmet}. From Table \ref{tab:sig_ilc_2lmet} we infer that to attain $5\sigma$ significance, we need 16 fb$^{-1}$, 23 fb$^{-1}$, 18 fb$^{-1}$ and 570 fb$^{-1}$ integrated luminosity ($\mathcal{L}_{5\sigma})$ for BP1, BP2, BP3 and BP4 respectively.

\begin{table}[ht!]
	\centering
		\resizebox{12cm}{!}{
	\begin{tabular}{|p{3.0cm}|c|c|c|p{3.0cm}|}
		\cline{2-4}
		\multicolumn{1}{c|}{}& \multicolumn{3}{|c|}{Number of Events after cuts ($\mathcal{L}=100$ fb$^{-1}$)} & \multicolumn{1}{c}{} \\ \cline{1-4}
		SM-background  
		 & $A_1$  &  $ A_2 $    &  $A_3$    & \multicolumn{1}{c}{}
		\\ \cline{1-4} 
		            $2 \ell + \mET$  & 4826  & 2817  & 535 (289)  \\ \cline{1-4} \hline
			\multicolumn{1}{|c|}{Signal }  &\multicolumn{3}{|c|}{} &\multicolumn{1}{c|}{$\mathcal{L}_{5\sigma}$ (fb$^{-1}$)}  \\ \cline{1-5}
		\multicolumn{1}{|c|}{BP1}  & 417 & 366 & 318 &\multicolumn{1}{|c|}{16}  \\ \hline
		\multicolumn{1}{|c|}{BP2} & 322  & 261 & 260 &  \multicolumn{1}{|c|}{23} \\ \hline 
		\multicolumn{1}{|c|}{BP3} &  291  &  225 & 223 & \multicolumn{1}{|c|}{18}  \\ \hline
		\multicolumn{1}{|c|}{BP4} &  45  & 36 & 36 & 
		 \multicolumn{1}{|c|}{570}  \\ \hline

	\end{tabular}}

	\caption{ The cut-flow for signal and backgrounds for $2 \ell + \mET$ channel along with the required integrated luminosity required for 5$\sigma$ significance for benchmarks 
	BP1, BP2, BP3 and BP4 at 1 TeV ILC. The bracketed term in the $A_3$ cut denotes the surviving number of events for $M_{\rm eff} < $ 350 GeV cut for the background.}
	\label{tab:sig_ilc_2lmet}
\end{table}

To showcase further improvement of the signal sensitivity from the cut-based analysis, we carry out the multivariate analysis (MVA) using {\em Decorrelated Boosted Decision Tree (BDTD)} algorithm within the Toolkit for Multivariate Data Analysis (TMVA) framework. A detailed description of the method has already been described in one of our earlier work \cite{Chakraborty:2021tdo}. According to the discerning ability between the signal and the backgrounds of this channel, the most important kinematic variables turn out to be \footnote{This is clearly in accordance with the variables highlighted in the cut-based analysis.} : 
\bea
M_{\ell_1 \ell_2},~ \eta_{\ell_1}, ~ \eta_{\ell_2}, ~ \mET, M_{\rm eff}, ~ p_T^{\ell_1} \,.
\eea

\begin{table}[htpb!]
\begin{center}
\resizebox{16cm}{!}{
\begin{tabular}{|c|c|c|c|c|c|}
\hline
 &  \hspace{5mm} {\texttt{NTrees}} \hspace{5mm} & \hspace{5mm} {\texttt{MinNodeSize}} \hspace{5mm} & \hspace{5mm} {\texttt{MaxDepth}}~~ \hspace{5mm} & \hspace{5mm} {\texttt{nCuts}} ~~\hspace{5mm} & \hspace{5mm} {\texttt{KS-score for}}~~\hspace{5mm}\\
 & & & & & {\texttt{Signal(Background)}} \\
\hline
\hline
\hspace{5mm} BP1 \hspace{5mm} & 110 & 4 \% & 2.0 & 40 & 0.545~(0.437) \\ \hline
\hspace{5mm} BP2 \hspace{5mm} & 110 & 4 \% & 2.0 & 50 & 0.303~(0.418) \\ \hline
\hspace{5mm} BP3 \hspace{5mm} & 110 & 3 \% & 2.0 & 50 & 0.969~(0.053) \\ \hline
\hspace{5mm} BP4 \hspace{5mm} & 110 & 3 \% & 2.0 & 50 & 0.018~(0.035) \\ \hline
\end{tabular}}
\end{center}
\caption{Tuned BDT parameters for BP1, BP2, BP3 and BP4 for the $2 \ell + \mET$ channel.}
\label{BDT-param-2lmet}
\end{table}

\begin{center}
\begin{table}[htb!]
\resizebox{17cm}{!}{
\begin{tabular}{|c|c|c|c|c|}\hline
 Benchmark Point & Signal Yield & Background Yield & Significance at 100 fb$^{-1}$ & $\mathcal{L}_{5\sigma}$ (fb$^{-1}$) \\ 
 & at 100 fb$^{-1}$  &at 100 fb$^{-1}$ &with 0\%(5\%) systematic uncertainty& with 0\%(5\%) systematic uncertainty \\ \hline 
BP1 & 346 & 711 & 12.1~(6.9)& 17.1~(52.5) \\ \hline
BP2 & 283  & 643 & 10.5~(6.7) & 22.8~(55.7) \\ \hline
BP3 & 214 &  203 & 13.1~(10.2) & 14.5~(24.0) \\ \hline
BP4 & 50 &  200 & 3.4~(2.7) & 216.3~(342.9) \\ \hline 
 \end{tabular}}
\caption{The signal and background yields at 1 TeV ILC with 100 fb$^{-1}$ integrated luminosity for BP1,BP2, 
BP3 and BP4 along with luminosity required for 5$\sigma$ significance
  for the $ e^+ e^- \rightarrow 2 \ell +\mET$ channel after performing 
the BDTD analysis. }
\label{BDTD-2lmet}
\end{table}
\end{center}
\begin{figure}[htpb!]{\centering
\subfigure[]{
\includegraphics[width=3in,height=2.45in]{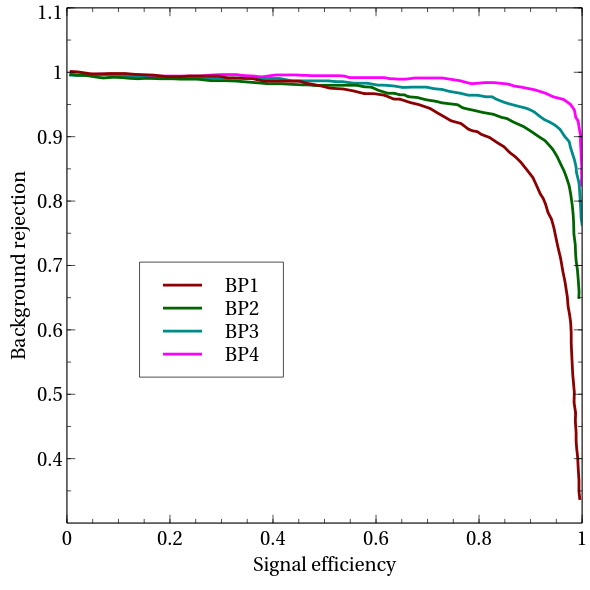}}
\subfigure[]{
\includegraphics[width=3.1in,height=2.46in]{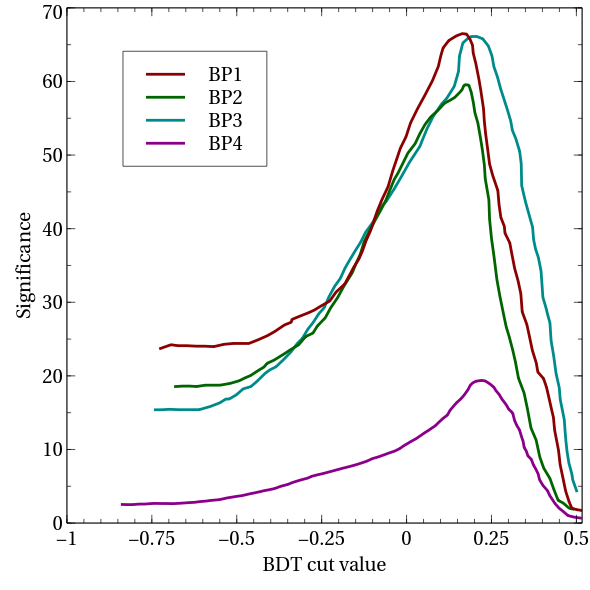}}}
\caption{ (a) ROC curves for chosen benchmark points for $2 \ell + \mET$ channel. (b) BDT-scores corresponding 
to BP1, BP2, BP3 and BP4 for $2 \ell + \mET$ channel.}
\label{ROC-BDTScore-2lmet}
\end{figure}
The BDTD parameters like \texttt{NTrees, MinNodeSize, MaxDepth, nCuts} and \texttt{KS-scores} \cite{Chakraborty:2021tdo} for both signal and backgrounds are tabulated in Table \ref{BDT-param-2lmet}. The first four  input parameters are regulated in such a way, that the \texttt{KS-scores} for both signal and backgrounds  become stable \cite{Chakraborty:2021tdo}. The next task is to tune the {\em BDT cut value } or {\em BDT score} to maximise the signal significance. Fig. \ref{ROC-BDTScore-2lmet}(b) shows the variation of significance with BDT-score. From this figure, it is evident that the signal significances of different benchmarks attain a maximum value for different BDT cut values. The BDT scores for BP1, BP2, BP3 and BP4 are 0.167, 0.164, 0.23 and 0.201 respectively. In the {\em Receiver's Operative Characteristic} (ROC) plot (Fig.\ref{ROC-BDTScore-2lmet}(a)), we show the degree of background rejection against signal efficiency. It can clearly be inferred that the degree of background rejection is maximum for BP4 (magenta curve in Fig.\ref{ROC-BDTScore-2lmet}(a)).

The signal and background yields for $2 \ell + \mET$ channel at $\cal{L}=$ 100 fb$^{-1}$ along with the integrated luminosity required to achieve a 5$\sigma$ significance for each benchmark points using MVA, have been tabulated in Table \ref{BDTD-2lmet}. The integrated luminosities required for achieving 5$\sigma$ significance for BP1, BP2, BP3, BP4 are 17.1, 22.8, 14.5, 216.3 respectively. Comparing with the results obtained from the cut-based analysis, one can find that the integrated luminosities required for achieving 5$\sigma$ significance is lowered for BP3 and BP4 after performing the BDTD analysis, which implies an overall improvement of results after the multivariate analysis is done. It is instructive to acknowledge systematic uncertainties at the 
experiment which can effect our results. To show this we include a $5\%$ systematic uncertainty and highlight its effect 
in Table \ref{BDTD-2lmet} along side the null systematic uncertainty results. The signal significance gets modified by 
introducing a systematic uncertainty ($\sigma_{sys\_un}$) in the SM background estimation~\cite{Adhikary:2020cli} following:
\begin{equation}
S_\text{sys} = \sqrt{2 \left((N_S+N_B) \log \left(\frac{(N_S+N_B)(N_B+\sigma_{B}^{2})}{N_B^{2}+(N_S+N_B)\sigma_{B}^{2}}\right)-\frac{N_B^{2}}{\sigma_{B}^{2}}\log \left(1+\frac{\sigma_{B}^{2}N_S}{N_B(N_B+\sigma_{B}^{2})} \right) \right)} \,,
\label{S_sys}
\end{equation}
where $\sigma_{B}=\sigma_{sys\_un}\times N_B$. 



\subsection{$1 \ell + 2j + \met $ final state}
The dominant contribution to $1 \ell + 2j + \mET $ final state originates from $ ~e^+ e^- \to E_1^+ E_1^- \to W^{+} W^{-} \mET$, where one of the $W^{\pm}$ decays leptonically and other one decays hadronically. The SM processes that can give rise to the similar final state are: \\
\begin{itemize}
\item $~e^+ e^- \rightarrow W^+W^-; ~ W^+(W^-) \to \ell^+(\ell^-)~ \nu_\ell(\bar{\nu}_\ell),~ W^-(W^+) \to j j$,
\item $~e^+ e^- \rightarrow ZZ; ~ Z \to \ell^+ \ell^-,~ Z \to j j $, (one of the leptons is missed)
\item $~e^+ e^- \rightarrow W^+W^-Z$;
\begin{enumerate}
\item $W^+ \to \ell^+ ~ \nu_\ell,~ ,~ W^- \to \ell^- ~ \bar{\nu}_\ell, ~ Z \to j j$, (one of the leptons is missed)
\item $W^+ \to \ell^+ ~ \nu_\ell,~ ,~ W^- \to j j, ~ Z \to \nu_\ell~ \bar{\nu}_\ell$,
 \end{enumerate}
\item $~ e^+ e^- \rightarrow ZZZ;~ Z \to \ell^+ \ell^-,~ Z \to j j , ~Z \to \nu_\ell ~ \bar{\nu}_\ell$, (one of the leptons is missed)
\item $ ~e^+ e^- \rightarrow Zh$; 
\begin{enumerate}
\item $Z \to \ell^+ \ell^-, ~ h \to j j$ (one of the leptons is missed)
\item $Z \to j j, ~ h \to \ell^+ \ell^-$ (one of the leptons is missed)
\end{enumerate}
\end{itemize}
Contributions from $ZZZ$ and $Zh$ backgrounds are insignificant due to small production rate. The LO cross sections of the signal and backgrounds using polarised and unpolarised incoming beams are tabulated in Table \ref{tab:bp_ilc_1l2j}.
\begin{table}[!hptb]
\centering
\resizebox{14cm}{!}{
 \begin{tabular}{|c|c|c|c|}
  \hline \hline
  & Cross section for  &  Cross section for   & Cross section for \\
  & ($P_{e^-},P_{e^+} = 0,0)$ & ($P_{e^-},P_{e^+} = 80\%L,30\%R)$ &  ($P_{e^-},P_{e^+} = 80\%R,0)$  \\
  & (in fb) &  (in fb) & (in fb) \\
  \hline
  \multicolumn{4}{|c|}{Signal benchmarks}  \\ \hline \hline
  BP1 & 19.30 & 33.39 & 14.50 \\ \hline
  BP2 & 17.27 & 31.04 & 11.76 \\ \hline
  BP3 & 9.11 & 16.36 & 6.20 \\ \hline
  BP4 & 1.65 & 3.0 & 1.10 \\ \hline
 \multicolumn{4}{|c|}{Background subprocesses  }  \\ \hline \hline
  $e^+ e^- \rightarrow W^+W^- $ & 229.21 & 535.74 & 33.08 \\ \hline
   $e^+ e^- \rightarrow ZZ $ &  3.41 & 5.77 & 2.65 \\ \hline
   $e^+ e^- \rightarrow W^+W^-Z $ & 3.82 & 8.89 & 0.57 \\ \hline
   $e^+ e^- \rightarrow ZZZ  $ &  0.01 & 0.02 & 0.009 \\ \hline
   $e^+ e^- \rightarrow Zh $ &  0.18 & 0.27 & 0.27 \\ \hline
  \hline
 \end{tabular}}
	\caption{ The effective cross-sections of
	the signal and backgrounds for $1\ell+ 2j + \mET$ channel at LO at 1 TeV ILC using unpolarised and polarised incoming beams. }
	\label{tab:bp_ilc_1l2j}
 \end{table}
 
  \begin{figure}[htpb!]{\centering
\subfigure[]{
\includegraphics[height = 5.5 cm, width = 8 cm]{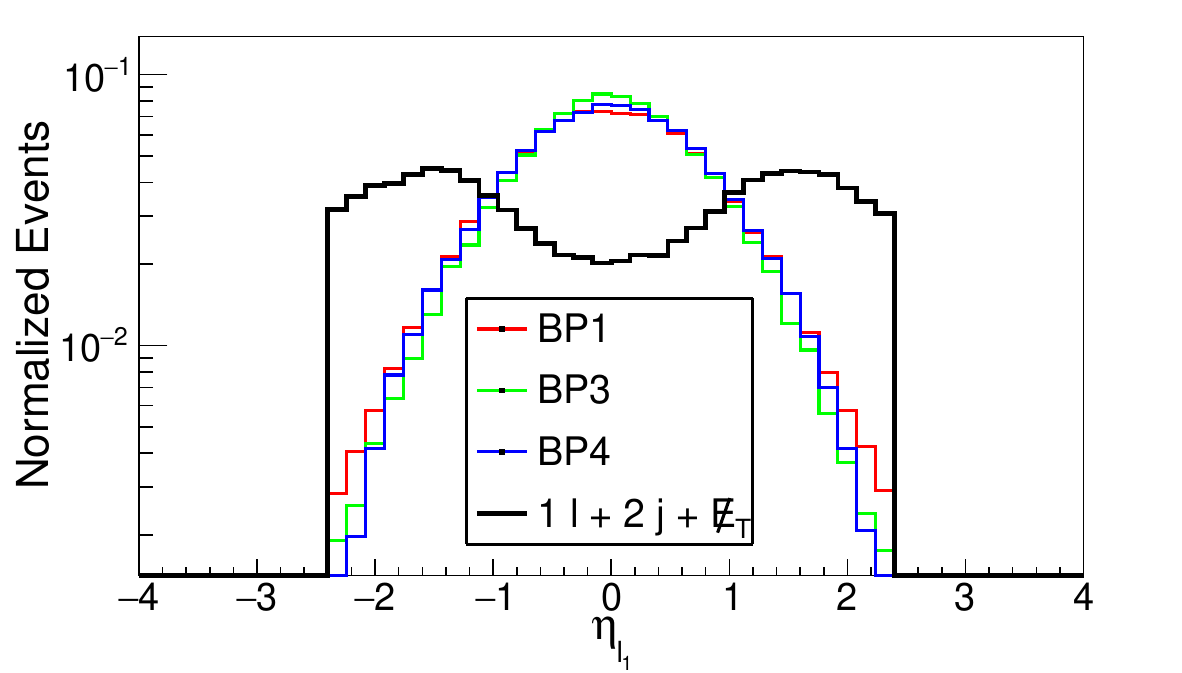}} \\
\subfigure[]{
\includegraphics[height = 5.5 cm, width = 8 cm]{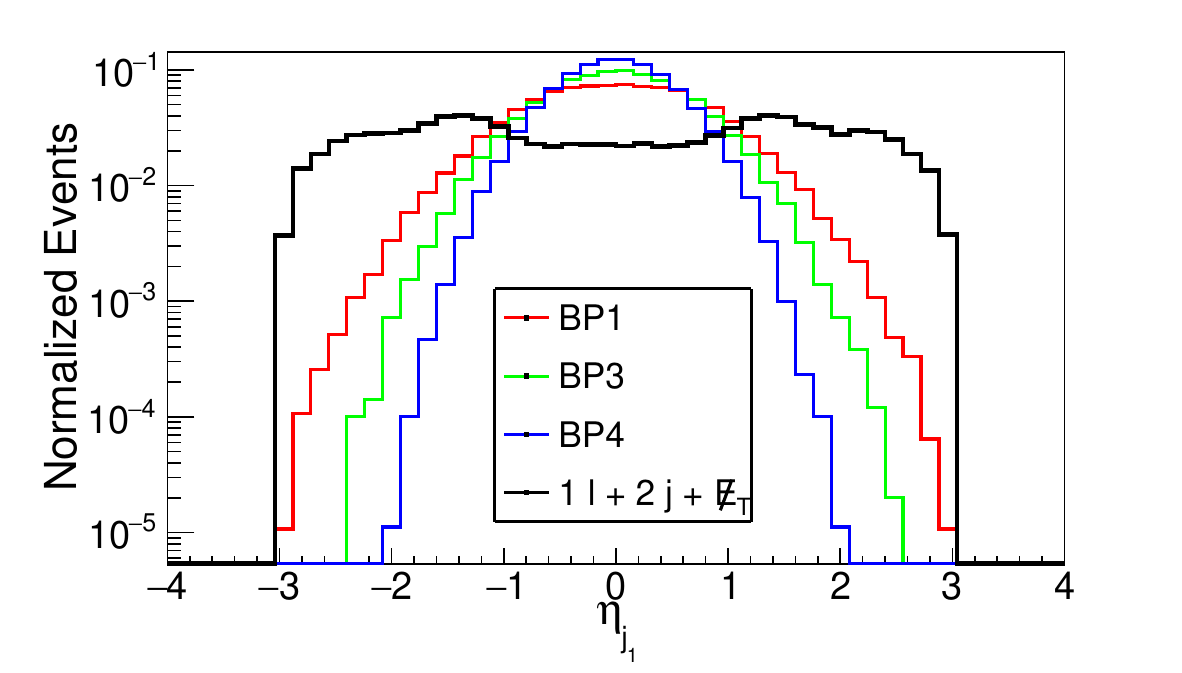}} 
\subfigure[]{
\includegraphics[height = 5.5 cm, width = 8 cm]{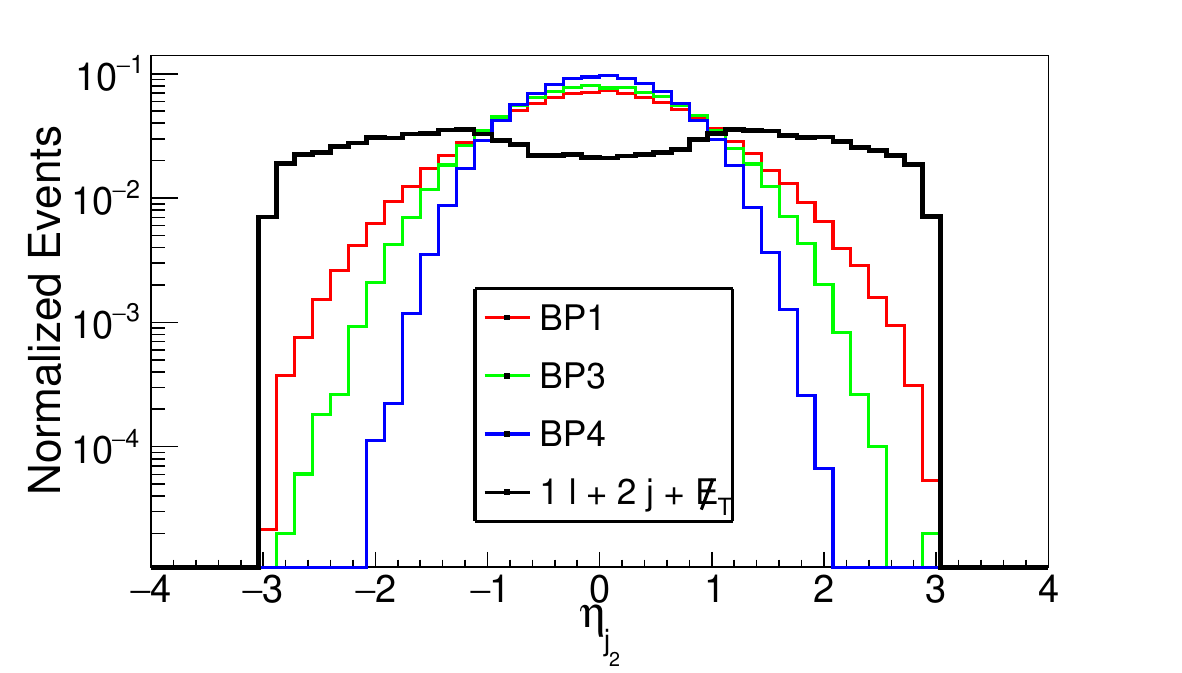}} \\
\subfigure[]{
\includegraphics[height = 5.5 cm, width = 8 cm]{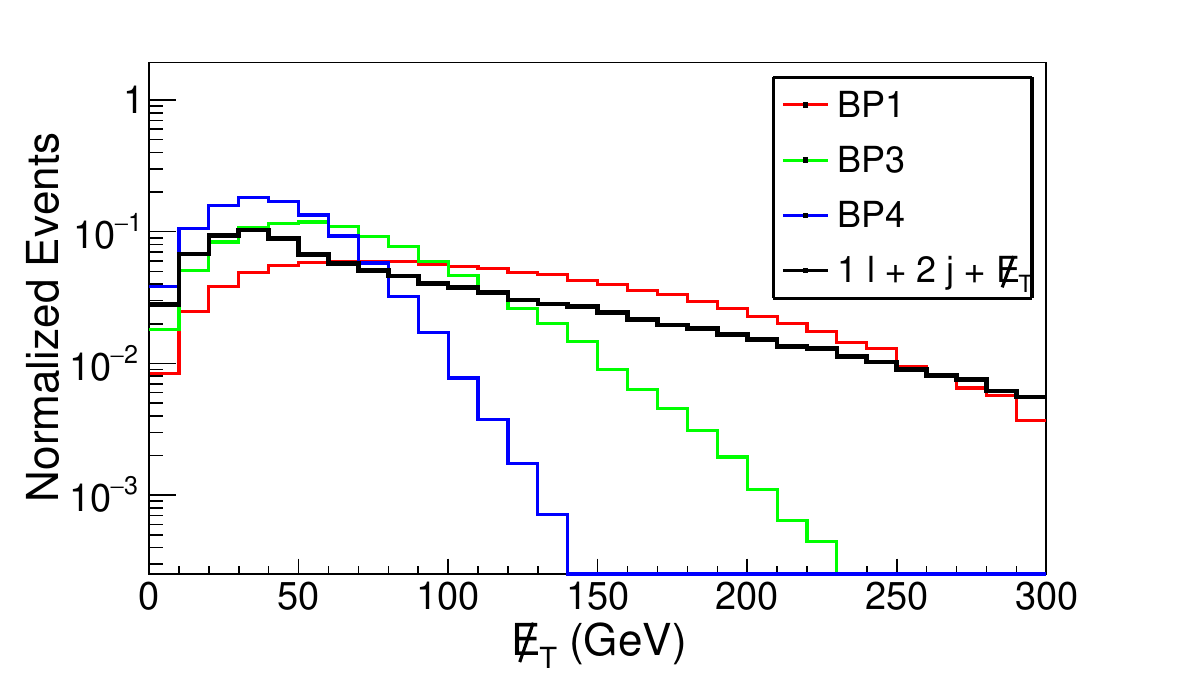}}
\subfigure[]{
\includegraphics[height = 5.5 cm, width = 8 cm]{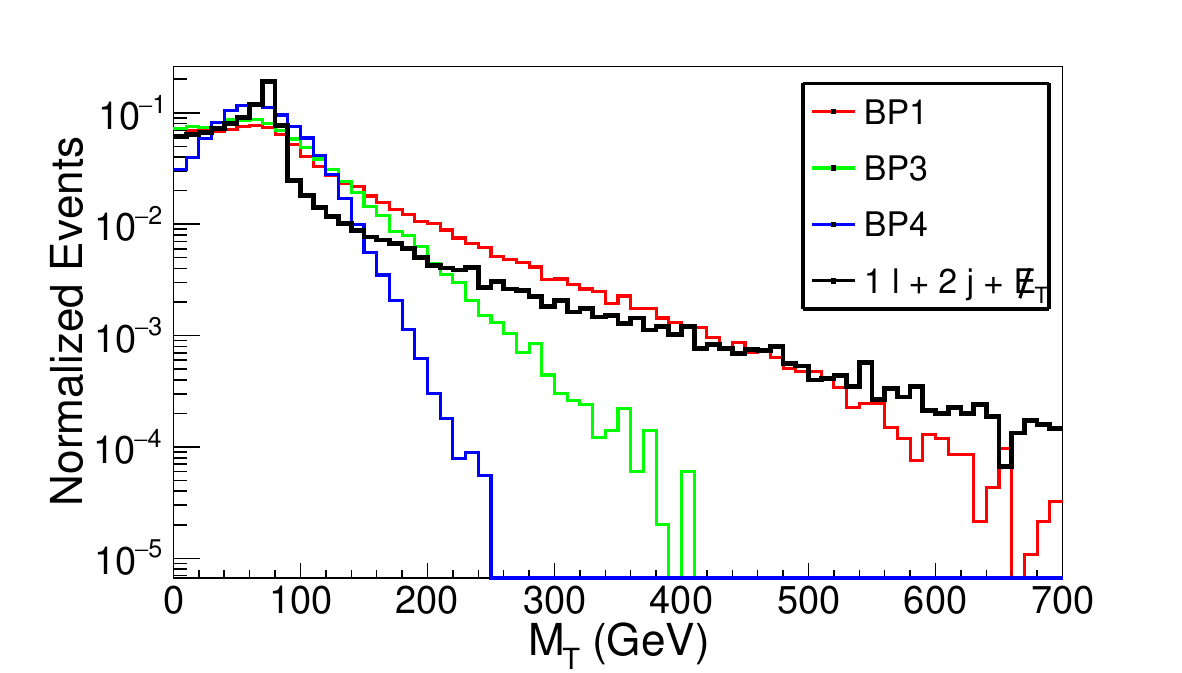}} 
}
\caption{ Normalized distributions of $ \eta_{\ell_1},~\eta_{j_1},~\eta_{j_2}, ~\mET, ~M_{T} $ for $1 \ell + 2j + \mET$ channel at 1 TeV ILC. }
\label{distribution_1l2j}
\end{figure}

\begin{figure}[htpb!]{\centering
\subfigure[]{
\includegraphics[height = 5.5 cm, width = 8 cm]{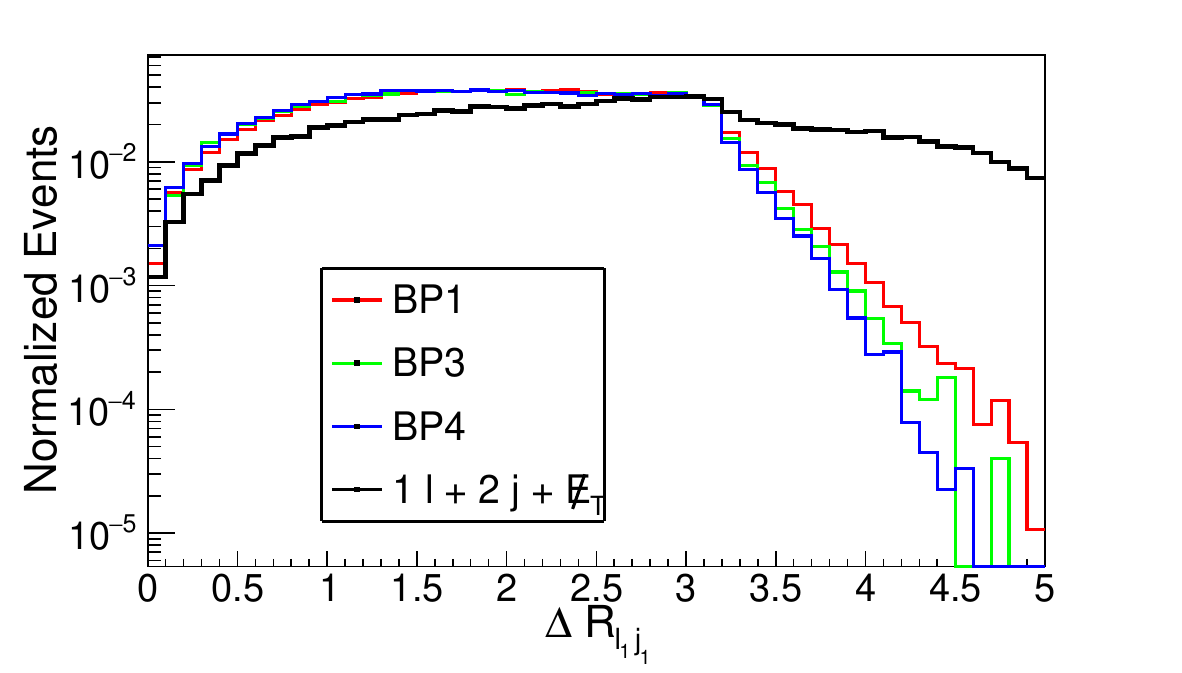}}
\subfigure[]{
\includegraphics[height = 5.5 cm, width = 8 cm]{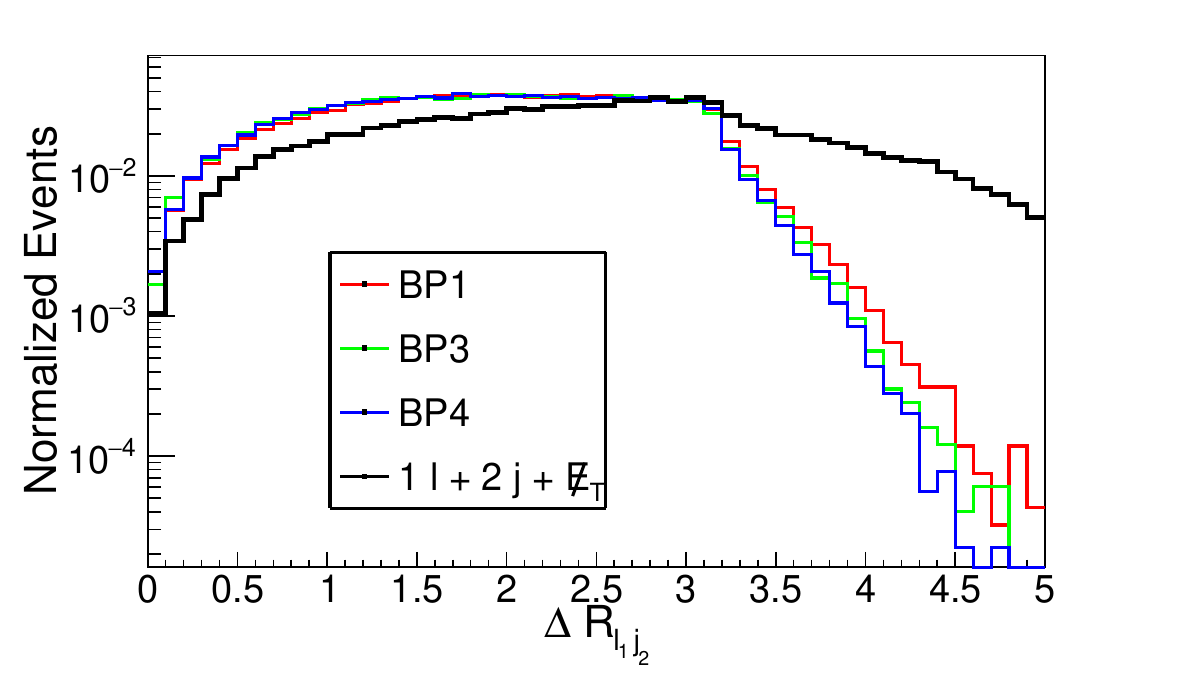}}
} 
\caption{ Normalized distributions of $~\Delta R_{\ell_1 j_{1}}, ~\Delta R_{\ell_1 j_{2}} $ for $1 \ell + 2j + \mET$ channel at 1 TeV ILC. }
\label{distribution_1l2j_2}
\end{figure}
 
 Since the signal consists of one lepton and two jets along with transverse missing energy, we reject any second lepton or any third jet in the final state for the backgrounds. This helps us to suppress $ZZ$ background. Finally we are left with $W^+W^-$ and $W^+W^-Z$ background. Apart from the basic acceptance cuts mentioned in Eq.(\ref{basic_cuts}), we implement the following cuts to enhance the signal over backgrounds.

 \begin{itemize}
 \item $B_1$: The pseudorapidity distributions for lepton and jets are different for the signal and background as can be seen in Fig~\ref{distribution_1l2j}(a),(b),(c) due to the t-channel dominant background. Choosing the pseudo-rapidity of the lepton and jets within the range : $|\eta_{\ell_1}|, |\eta_{j_{1,2}}| < 1.0$, helps to reduce the background drastically. 
 
  \item $B_2$: The normalized $\mET$ distribution is shown in  Fig~\ref{distribution_1l2j}(d).  For the background, the missing energy comes from the neutrinos and the distribution peaks at a lower $\mET$ value. On the other hand, for the signal, apart from the neutrinos, missing energy can arise from the DM candidates and as a result it will peak relatively at a higher value. A lower cut of $\mET > 50$ GeV helps to diminish the background.
  
  \item $B_3$: We use the kinematic variable transverse mass $M_T$ \footnote{$M_T$ is defined as $M_T= \sqrt{2 p^{\ell}_T \mET ~(1- \cos~ \Delta \phi_{\ell,\mET}~)}$,  where $\Delta \phi_{\ell,\mET}$ is the azimuthal angle between the lepton and transverse missing energy. } to distinguish the signal and background. As expected for the background, $M_T$ will peak at the $W^{\pm}$ mass while for the signal the corresponding distribution is smeared as seen in Fig~\ref{distribution_1l2j}(e). As there is additional source of $\mET$ in the signal, we get a tail in $M_T$ distribution for the signal. We observe that putting a cut of $M_T > 90$ GeV helps to suppress the background. 
 
  \item $B_4$: We depict the $\Delta R_{\ell_1 j_{1,2}}$ distributions in Fig~\ref{distribution_1l2j_2}(a),(b). It can be seen that an upper cut of $\Delta R_{\ell_1 j_{1,2}} < 3.2$ helps to enhance the signal significance. 
 \end{itemize}
 
 We show the effect of the each cut in Table~\ref{tab:1l2j_cut}. It is noticed that after putting all the cuts, we merely need 4, 3 and 14 fb$^{-1}$ integrated luminosity to probe BP1, BP2 and BP3 for achieving 5$\sigma$ significance. However, due to small production cross-section, to probe BP4 we need comparatively higher (220 fb$^{-1}$) luminosity.  
 
\begin{table}[ht!]
	\centering
		\resizebox{12cm}{!}{
	\begin{tabular}{|p{3.0cm}|c|c|c|c|p{3.0cm}|}
		\cline{2-5}
		\multicolumn{1}{c|}{}& \multicolumn{4}{|c|}{Number of Events after cuts ($\mathcal{L}=100$ fb$^{-1}$)} &  \multicolumn{1}{c}{} \\ \cline{1-5}
		SM-background  
		 & $B_1$  &  $ B_2 $    &  $B_3$ & $B_4$   & \multicolumn{1}{c}{}
		\\ \cline{1-5} 
           $W^+W^-$ & 1008  & 873 & 39 & 26 \\ \cline{1-5}
           $W^+W^-Z$ & 101 & 85 & 33 & 28  
          
		  \\ \hline \hline
		
			\multicolumn{1}{|c|}{Signal }  &\multicolumn{4}{|c|}{} &  \multicolumn{1}{|c|}{$\mathcal{L}_{5\sigma}$ (fb$^{-1}$)} \\ \cline{1-6} 
		\multicolumn{1}{|c|}{BP1}    & 1129  & 891 &  329 & 310 & \multicolumn{1}{|c|}{4}   \\ \hline
		\multicolumn{1}{|c|}{BP2} & 915   & 710 & 330 & 314  & \multicolumn{1}{|c|}{3} \\ \hline 
		\multicolumn{1}{|c|}{BP3} & 526   & 334 & 134 &  127 & \multicolumn{1}{|c|}{14} \\ \hline 
		\multicolumn{1}{|c|}{BP4} & 108   & 42  & 28 &  27 & \multicolumn{1}{|c|}{220} \\ \hline

	\end{tabular}}

	\caption{ The cut-flow for signal and backgrounds along with the significances for BP1, BP2, BP3 and BP4 at 1 TeV ILC and the required integrated luminosity for 5$\sigma$ significance for the $ e^+ e^- \rightarrow 1 \ell+2j+\mET$ channel. }
	\label{tab:1l2j_cut}
\end{table}		

\begin{table}[htpb!]
\begin{center}
\resizebox{16cm}{!}{
\begin{tabular}{|c|c|c|c|c|c|}
\hline
 &  \hspace{5mm} {\texttt{NTrees}} \hspace{5mm} & \hspace{5mm} {\texttt{MinNodeSize}} \hspace{5mm} & \hspace{5mm} {\texttt{MaxDepth}}~~ \hspace{5mm} & \hspace{5mm} {\texttt{nCuts}} ~~\hspace{5mm} & \hspace{5mm} {\texttt{KS-score for}}~~\hspace{5mm}\\
 & & & & & {\texttt{Signal(Background)}} \\
\hline
\hline
\hspace{5mm} BP1 \hspace{5mm} & 110 & 3 \% & 2.0 & 50 & 0.263~(0.048) \\ \hline
\hspace{5mm} BP2 \hspace{5mm} & 110 & 3 \% & 2.0 & 50 & 0.576~(0.242) \\ \hline
\hspace{5mm} BP3 \hspace{5mm} & 110 & 4 \% & 2.0 & 50 & 0.071~(0.197) \\ \hline
\hspace{5mm} BP4 \hspace{5mm} & 110 & 3 \% & 2.0 & 50 & 0.809~(0.307) \\ \hline
\end{tabular}}
\end{center}
\caption{Tuned BDT parameters for BP1, BP2, BP3 and BP4 for the $1 \ell + 2 j + \mET$ channel.}
\label{BDT-param-1l2jmet}
\end{table}

\begin{center}
\begin{table}[htb!]
\resizebox{18cm}{!}{
\begin{tabular}{|c|c|c|c|c|}\hline
 Benchmark Point & Signal Yield & Background Yield & Significance at 100 fb$^{-1}$ & $\mathcal{L}_{5\sigma}$ (fb$^{-1}$) \\ 
 & at 100 fb$^{-1}$  &at 100 fb$^{-1}$ &with 0\%(5\%) systematic uncertainty& with 0\%(5\%) systematic uncertainty \\ \hline 
BP1 & 1531 & 641  & 47.3~(25.4) &1.1~(3.9) \\ \hline
BP2 & 1480  & 301 & 58.0~(40.3) &0.7~(1.5) \\ \hline
BP3 & 834 &  224 & 40.2~(28.5) &1.5~(3.1) \\ \hline
BP4 & 151 &  41 & 17.0~(15.6)&8.6~(10.3) \\ \hline 
 \end{tabular}}
\caption{The signal and background yields at 1 TeV ILC with 100 fb$^{-1}$ integrated luminosity for BP1,BP2, 
BP3 and BP4 along with luminosity required for  5$\sigma$ significance for the $ e^+ e^- \rightarrow 1 \ell + 2 j +\mET$ channel after performing 
the BDTD analysis. }
\label{BDTD-1l2jmet}
\end{table}
\end{center}
\begin{figure}[htpb!]{\centering
\subfigure[]{
\includegraphics[width=3in,height=2.45in]{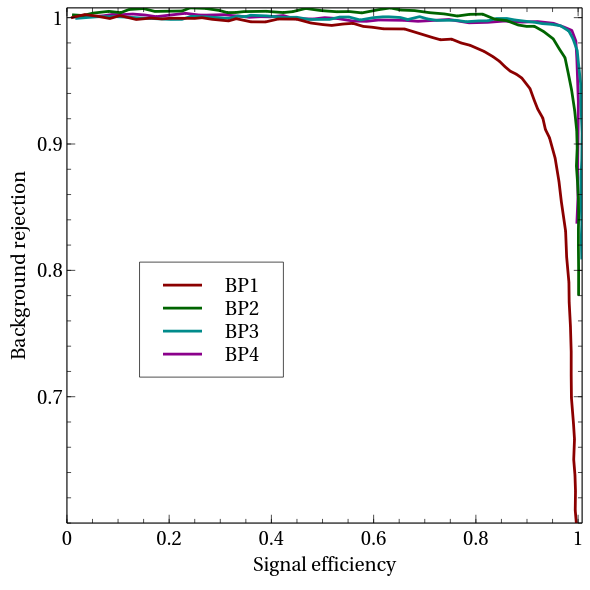}}
\subfigure[]{
\includegraphics[width=3.1in,height=2.46in]{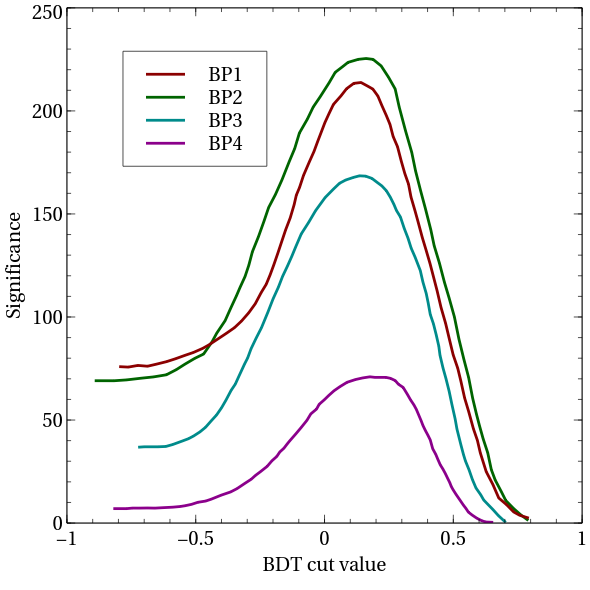}}}
\caption{ (a) ROC curves for chosen benchmark points for $1 \ell + 2 j + \mET$ channel. (b) BDT-scores corresponding 
to BP1, BP2, BP3 and BP4 for $1 \ell + 2 j + \mET$ channel.}
\label{ROC-BDTScore-1l2jmet}
\end{figure}

We now perform the multivariate analysis for $1 \ell + 2 j + \mET$ channel. According to the degree of differentiating potential between the signal and backgrounds, the most important variables turn out to be :
\bea
 \Delta R _{\ell_1 j_1}, ~ \Delta R _{\ell_1 j_2}, ~ M_T,~ \mET, ~ \eta_{\ell_1},~ \eta_{j_1}, ~\eta_{j_2},~ M_{j_1 j_2},~ \Delta \phi_{\ell_1 j_1}, ~ \Delta \phi_{\ell_1 j_2}
 \eea
  Here $\Delta \phi_{\ell_1 j_1}, ~ (\Delta \phi_{\ell_1 j_2})$ are the azimuthal angle between $\ell_1$ and $j_1~(j_2)$, while the other variables have been defined earlier. The tuned BDT parameters for each benchmark points are listed in Table \ref{BDT-param-1l2jmet}. The signal and background yields for an integrated luminosity 100 fb$^{-1}$ are shown in Table \ref{BDTD-1l2jmet}. The same table contains the necessary integrated luminosities to attain 5$\sigma$ significance for all benchmarks. Fig.\ref{ROC-BDTScore-1l2jmet}(a) and \ref{ROC-BDTScore-1l2jmet}(b) depict the ROC curves and variation of significances with BDT-scores for all benchmarks respectively. The BDT scores for the four benchmarks are 0.13,0.165, 0.193, 0.141 respectively.


\subsection{$4\ell + \met $ final state}
In this section, we analyse the final state comprising $4 \ell + \mET$.
The $4\ell+\mET$ final state for the signal can be obtained from the following processes:
\begin{equation}
 e^+ e^- \to N_i N_i, N_i \to N_1 \ell^{+} \ell^{-}, ~~{\rm with} ~~i = 2,3,..8.
\end{equation}

 The SM backgrounds~\cite{Aaboud:2018zeb} that give rise to the 
similar final state is $VVV,(V= W^{\pm},Z$) production along with additional contribution coming from $ZZ$ production. Demanding that opposite sign same flavor (OSSF) lepton pair invariant mass lies away from the $Z$ peak reduces the $ZZ \to 4\ell$ and $ZZZ$ background significantly. Finally we are left with the irreducible $W^+W^-Z$ background. The signal and background cross sections at LO are depicted in Table \ref{tab:bp_ilc_4l}.

\begin{table}[!hptb]
\centering
\resizebox{14cm}{!}{
 \begin{tabular}{|c|c|c|c|}
  \hline \hline
  & Cross section for  &  Cross section for   & Cross section for \\
  & ($P_{e^-},P_{e^+} = 0,0)$ & ($P_{e^-},P_{e^+} = 80\%L,30\%R)$ &  ($P_{e^-},P_{e^+} = 80\%R,0)$  \\
  & (in fb) &  (in fb) & (in fb) \\
  \hline
  \multicolumn{4}{|c|}{Signal benchmarks}  \\ \hline \hline
  BP1 & 0.029 & 0.05 & 0.02 \\ \hline
  BP2 & 0.056 & 0.1 & 0.038 \\ \hline
  BP3 & 0.006 & 0.01 & 0.004 \\ \hline
  BP4 & 0.001 & 0.002 & 0.0007\\ \hline
  \multicolumn{4}{|c|}{Background }  \\ \hline \hline
  $e^+ e^- \rightarrow W^+W^-Z$ & 0.19 & 0.44 & 0.029 \\ \hline
  \hline
 \end{tabular}}
	\caption{ The effective cross-sections of
	the signal and background for $4\ell+ \mET$ signal at LO at 1 TeV ILC using unpolarised and polarised incoming beams. }
	\label{tab:bp_ilc_4l}
 \end{table}
 
 Along with the basic cuts (Eq.\ref{basic_cuts}), we implement the following cuts to maximise the signal significance :
 \begin{itemize}
  \item $C_1$: Out of the four leptons, we choose two pairs of OS same flavor leptons ($(M_{\ell^+ \ell^-})_{1,2}$ ) which have invariant mass close to the $Z$-mass. We reject all events where $ |(M_{\ell^+ \ell^-})_{1,2} - M_Z| < 15$ GeV to exclude the 
 $Z$-peak of $ZZ$-background.
  \item $C_2$: The pseudo-rapidity distributions of the leading and sub-leading leptons look similar to Fig.~\ref{distribution-dilep}(a),(b) as the t-channel contribution dominates. The $ |\eta_{\ell_{1,2}}| < 1.0$ cut helps to suppress the background.
  \item $C_3$: The $\mET$ distribution for the background peaks at a lower value as it gets contribution only from neutrinos unlike the signal that also gets contribution from the heavy dark matter. A lower cut on $\mET > 30$ GeV helps to enhance the signal significance.
  \item $C_4$: The normalized $M_{\rm eff}$ distribution is similar to Fig.~\ref{distribution-dilep}(d), except a larger tail. This is due to the fact that instead of two leptons, here $M_{\rm eff}$ includes the scalar sum of four lepton $p_T$'s and the missing transverse energy. We have optimized $M_{\rm eff} < 600 (500)$ GeV for BP1(rest of the BPs) to suppress the background significantly.
 \end{itemize}

 We tabulate the surviving events for the signal and backgrounds after each cut in Table~\ref{tab:fourlep_cut} at an integrated luminosity 4 ab$^{-1}$. It can be seen that to probe benchmark BP1 and BP2 at $5\sigma$ significance, we need 550 fb$^{-1})$ and 150 fb$^{-1})$ luminosity and BP3 and BP4 are beyond the ILC projected luminosity~\cite{Behnke:2013xla,Baer:2013cma}.

 \begin{table}[ht!]
	\centering
		\resizebox{11.5cm}{!}{
	\begin{tabular}{|p{3.0cm}|c|c|c|c|p{3.0cm}|}
		\cline{2-5}
		\multicolumn{1}{c|}{}& \multicolumn{4}{|c|}{Number of Events after cuts ($\mathcal{L}=4$ ab$^{-1}$)} &  \multicolumn{1}{c}{} \\ \cline{1-5}
		SM-background  
		 & $C_1$  &  $ C_2 $    &  $C_3$ & $C_4$   & \multicolumn{1}{c}{}
		\\ \cline{1-5} 
           $W^+ W^- Z$ & 81  & 35  & 32 & 5(2)
		  \\ \hline \hline
			\multicolumn{1}{|c|}{Signal }  &\multicolumn{4}{|c|}{} &  \multicolumn{1}{|c|}{$\mathcal{L}_{5\sigma}$ (fb$^{-1}$)} \\ \cline{1-6} 
		\multicolumn{1}{|c|}{BP1}    & 95  & 69 & 63 & 55 & \multicolumn{1}{|c|}{550}   \\ \hline
		\multicolumn{1}{|c|}{BP2} &  188  & 132 & 110 & 109 & \multicolumn{1}{|c|}{150} \\ \hline
		 \multicolumn{1}{|c|}{BP3} & 26 & 19 & 10 & 10  & \multicolumn{1}{|c|}{4500} \\ \hline 
		\multicolumn{1}{|c|}{BP4} & 3 & 3 & 2 & 2 & \multicolumn{1}{|c|}{65000} \\ \hline 
       
	\end{tabular}}

	\caption{ The cut-flow for signal and backgrounds for BP1, BP2, BP3 and BP4 at 1 TeV ILC and the required integrated luminosity for 5$\sigma$ significance for the $ e^+ e^- \rightarrow 4 \ell+\mET$ channel. The bracketed term in $C_4$ cut denotes the surviving number of events for $M_{\rm eff} < $ 500 GeV cut for the background.}
	\label{tab:fourlep_cut}
\end{table}	
\subsection{$4 j+\mET$ final state}
\label{4jmet}
This final state originates from $e^+ e^- \to E_1^+ E_1^- \to W^{+} W^{-} \mET$ process, where both the $W^{\pm}$ decay hadronically. The background for this process comes from $e^+ e^- \to 4 j+ \mET$ which is dominated by di-boson (in that part of the phase space where $\mET$ measurement is not important) and tri-boson production. However, due to small cross-section, $ZZZ$ contributes insignificantly, while $W^{+}W^{-}$, $ZZ$ and $W^{+}W^{-}Z$ act as the irreducible backgrounds for this signal. We have demanded a $b$-veto to reduce the  $t\bar{t}$ background ($t\bar{t}$ production cross-section is one order of magnitude less than $ZZ$ production cross-section and two orders of magnitude less than $W^{+}W^{-}$ production cross-section) as the efficiency of mistagging a $b$ jet as light jet is 1$\%$. The effective cross-sections for the signal and backgrounds are shown in Table~\ref{tab:bp_ilc_4j}.

\begin{table}[!hptb]
\centering
\resizebox{13cm}{!}{
 \begin{tabular}{|c|c|c|c|}
  \hline \hline
 & Cross section for  &  Cross section for   & Cross section for \\
  & ($P_{e^-},P_{e^+} = 0,0)$ & ($P_{e^-},P_{e^+} = 80\%L,30\%R)$ &  ($P_{e^-},P_{e^+} = 80\%R,0)$  \\
  & (in fb) &  (in fb) & (in fb) \\
  \hline
  \multicolumn{4}{|c|}{Signal benchmarks}  \\ \hline \hline
  BP1 & 20.21 & 34.97 & 15.18 \\ \hline
  BP2 & 18.82 & 33.66 & 12.75 \\ \hline
  BP3 & 7.21 & 12.95 & 4.91 \\ \hline
  BP4 & 1.20 & 2.19 & 0.80 \\ \hline
  \multicolumn{4}{|c|}{Background}  \\ \hline \hline

   $e^+ e^- \rightarrow 4j+\mET$ & 267.69  & 594.57  & 77.69 \\ \hline
   
  \hline
 \end{tabular}}
	\caption{ The effective cross-sections of
	the signal and background for $4j + \mET$ channel at LO at 1 TeV ILC using unpolarised and polarised incoming beams. }
	\label{tab:bp_ilc_4j}
 \end{table}

 To ensure that our signal contains exactly four jets, we veto any fifth jet with $p_T^j > 20$ GeV along with the basic cuts described in Eq.(\ref{basic_cuts}). In addition to these cuts, we put the following set of cuts to suppress the background. 

\begin{figure}[htpb!]{\centering
\subfigure[]{
\includegraphics[height = 5.5 cm, width = 8 cm]{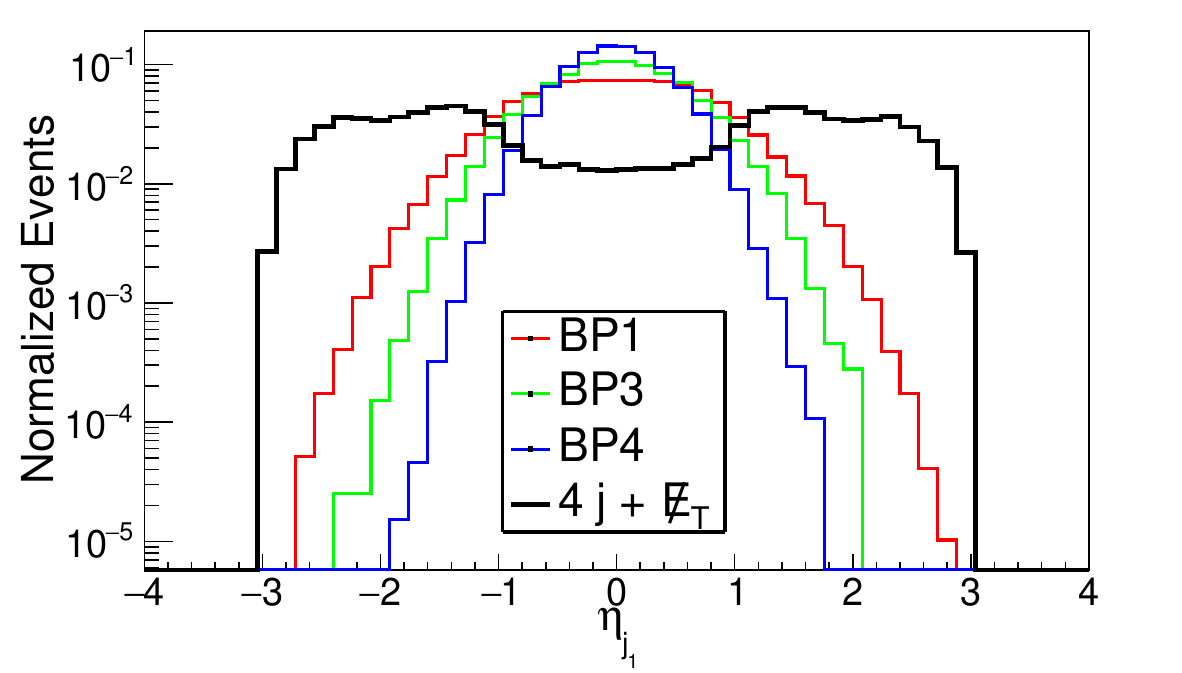}} 
\subfigure[]{
\includegraphics[height = 5.7 cm, width = 8.0 cm]{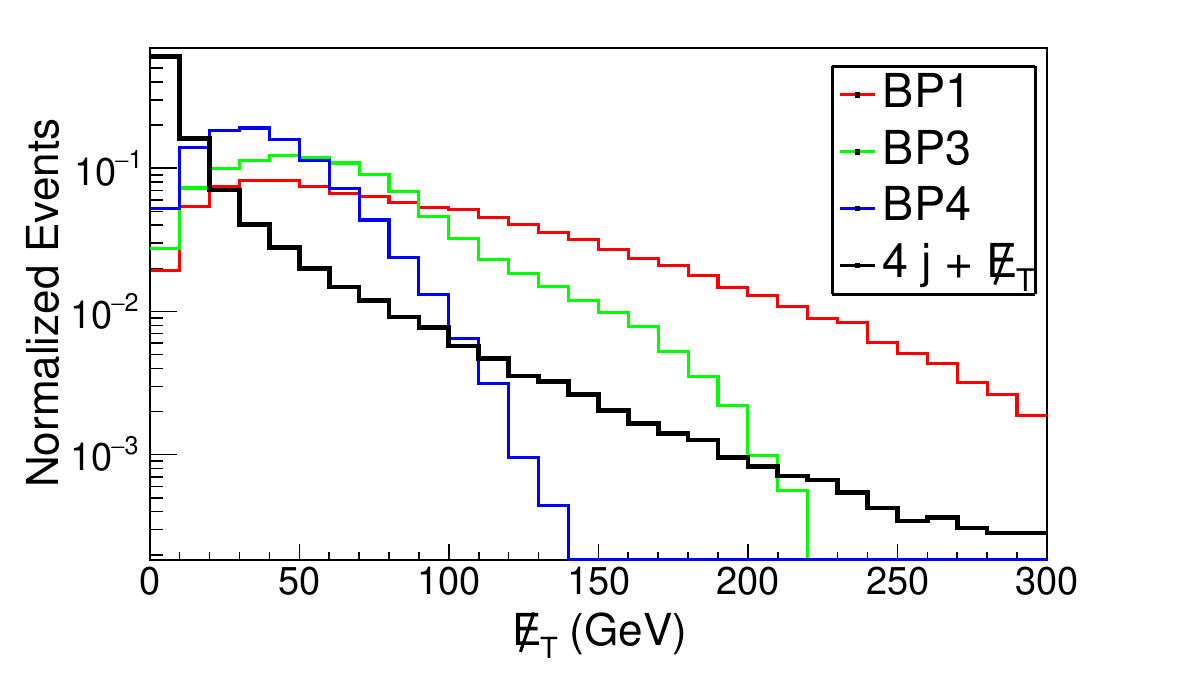}} \\
\subfigure[]{
\includegraphics[height = 5.5 cm, width = 8 cm]{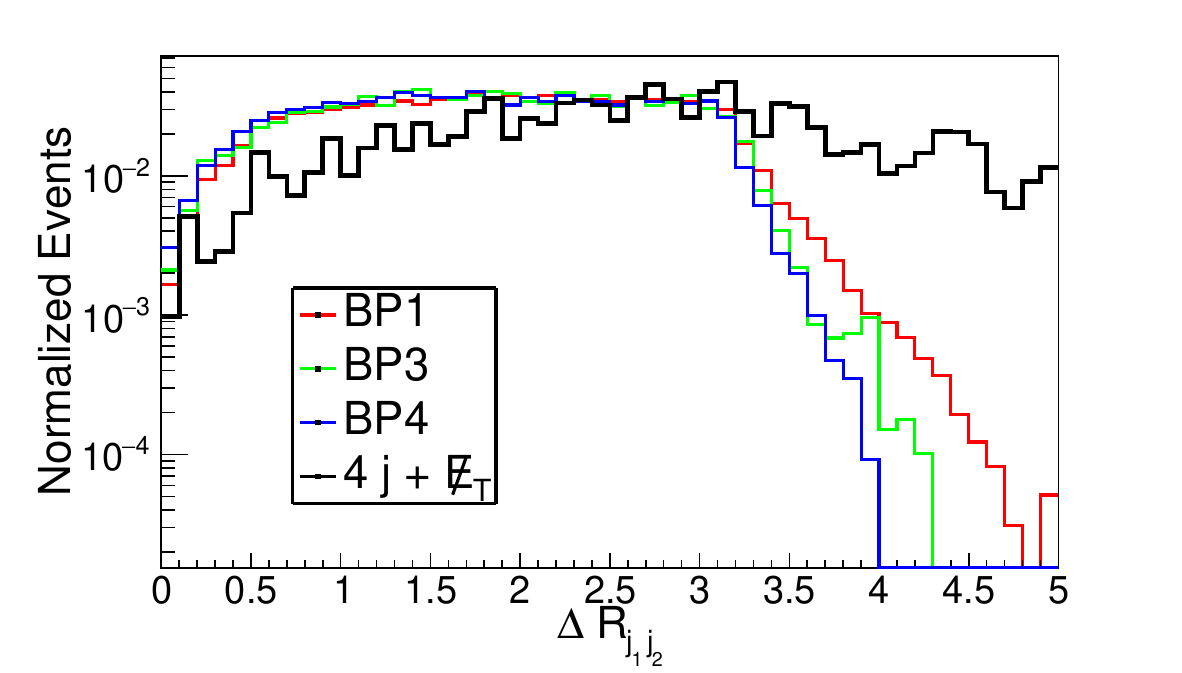}} 
\subfigure[]{
\includegraphics[height = 5.7 cm, width = 8.0 cm]{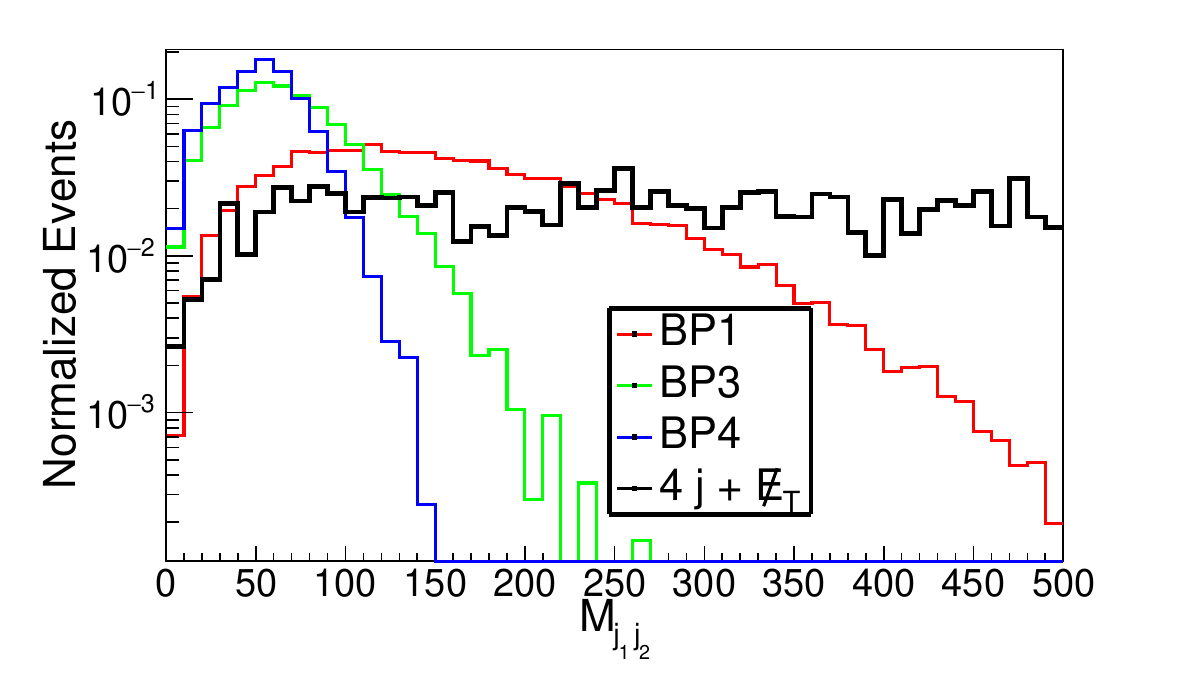}} 
}

\caption{ Normalized distributions of $ \eta_{j_1}, ~\mET, \Delta R_{j_{1} j_{2}}, ~M_{j_{1} j_{2}} $ for $4j + \mET$ channel at 1 TeV ILC.  }
\label{distribution_4j}
\end{figure}

\begin{itemize}
 \item $D_1$: We draw normalized pseudo-rapidity distribution in Fig.~\ref{distribution_4j}(a) for the leading jet.  
For the signal, the jets are much more centralised. Therefore putting a cut of $|\eta_j| < 1.2, j = 1...4$, helps to suppress the background very efficiently. 
 
 \item $D_2$: For the background, the source of $\mET$ is only the neutrinos coming from the decay of $W^{\pm}$ or $Z$. For the signal, the additional source is the massive dark matter. By examining the distribution as depicted in Fig.~\ref{distribution_4j}(b), we put a cut of $\mET > 30$ GeV to enhance the signal. 
 
 \item $D_3$: $\Delta R$ between the jets become an important variable. We put a cut of $\Delta R_{j_{i} j_{k}} < 3.5, i \neq k = 1...4$, to suppress the background. 
 
 \item $D_4$: The invariant mass for two jet pair becomes an efficient variable. For BP1 and BP2, since the mass difference between the charged and neutral VLL's is higher compared to BP3 and BP4, the invariant mass distribution has a larger tail. As seen from Fig.~\ref{distribution_4j}(d), $M_{j_{i}j_{k}} < 300(150)$ GeV, $i \neq k = 1...4$, helps to enhance the signal for benchmark BP1 and BP2 (BP3 and BP4) over the background.
 
\end{itemize}

We present the cut-flow for the signal and backgrounds at integrated luminosity 100 fb$^{-1}$ in Table~\ref{tab:sig_ilc_4jmet}. We observe that compared to all 
the remaining aforementioned channels, this channel performs the best.

\begin{table}[ht!]
	\centering
		\resizebox{17cm}{!}{
	\begin{tabular}{|p{3.0cm}|c|c|c|c|p{3.0cm}|p{3.0cm}|}
		\cline{2-5}
		\multicolumn{1}{c|}{}& \multicolumn{4}{|c|}{Number of Events after cuts ($\mathcal{L}=100$ fb$^{-1}$)} & \multicolumn{1}{c}{} \\ \cline{1-5}
		SM-background  
		 & $D_1$  &  $ D_2 $    &  $D_3$  & $D_4$  & \multicolumn{1}{c}{}
		\\ \cline{1-5} 
                             $4j+\mET$ & 3554 & 1198 & 326 & 134(27) \\ \cline{1-7}  
			\multicolumn{1}{|c|}{Signal }  &\multicolumn{4}{|c|}{}& \multicolumn{1}{|c|}{Significance at 100 fb$^{-1}$} &\multicolumn{1}{c|}{$\mathcal{L}_{5\sigma}$ (fb$^{-1}$)}  \\ 
			\multicolumn{1}{|c|}{ } & \multicolumn{4}{|c|}{} & \multicolumn{1}{|c|}{  with 0\%(5\%) uncertainty} & \multicolumn{1}{c|}{with 0\%(5\%) uncertainty} \\ \cline{1-7}
		\multicolumn{1}{|c|}{BP1}  & 1263 & 1096 & 1009 & 921 & ~~~~~~~~~~~~~~~~~~~50.1~(37.9) &\multicolumn{1}{|c|}{1.0~(1.7)}  \\ \hline
		\multicolumn{1}{|c|}{BP2} &  1224  & 1058 & 990 & 989 & ~~~~~~~~~~~~~~~~~~~52.8~(39.8) &\multicolumn{1}{|c|}{0.9~(1.6)} \\ \hline 
		\multicolumn{1}{|c|}{BP3} & 507   & 407   & 382 & 370 & ~~~~~~~~~~~~~~~~~~~37.3~(33.7) &\multicolumn{1}{|c|}{1.8~(2.2)}  \\ \hline
		\multicolumn{1}{|c|}{BP4} &  106  & 66 & 63 & 63 & ~~~~~~~~~~~~~~~~~~~9.5~(9.0)
		 &\multicolumn{1}{|c|}{27.7~(30.9)}  \\ \hline

	\end{tabular}}

	\caption{ The cut-flow for signal and backgrounds for $4 j + \mET$ channel along with the required integrated luminosity required for 5$\sigma$ significance for benchmarks 
	BP1, BP2, BP3 and BP4 at 1 TeV ILC. The bracketed term in the $D_4$ cut denotes the surviving number of events for $M_{j_{i}j_{k}} <  150$ GeV cut for the background. }
	\label{tab:sig_ilc_4jmet}
\end{table}

As we can achieve 5$\sigma$ significance with low integrated luminosity with cut-based analysis already, we refrain ourselves in performing the multivariate analysis.
Comparing with the previous channels, this channel fares the best among all at 1 TeV ILC .

We conclude this section by a thorough comparison of the four aforementioned channels. We have quoted the required luminosity to probe the benchmark points with 5$\sigma$ significance by cut-based analysis only. From Table~\ref{compare}, it can be seen that out of the four channels, $4j+\mET~$ channel performs the best as it requires $< 2$ fb$^{-1}$ luminosity to probe the first three BP's and 27.7 fb$^{-1}$ luminosity to probe BP4 with $5\sigma$ significance. $1\ell + 2j+\mET$ channel performs the second best to probe the selected benchmarks. The benchmark points BP1, BP2, BP3 and BP4 can be probed with $5\sigma$ significance with luminosity 4, 3, 14 and 220 fb$^{-1}$ respectively in this mode. The third best performing channel is  
$2\ell+\mET$. To probe the first three benchmark points with $5\sigma$ significance one requires luminosity $< 25$ fb$^{-1}$. However, due to small effective cross-section, probing the BP4 at $5\sigma$, one requires 570 fb$^{-1}$ luminosity. The $4\ell+\mET$ channel performs the worst at ILC although the background cross-sections are small. This is due to the fact that the effective production-cross section of signals in $4\ell+\mET$ channel is very small. Only for BP1 and BP2, $5\sigma$ significance can be achieved with 500 fb$^{-1}$ and 150 fb$^{-1}$ integrated luminosity respectively. Thus one can conclude that, the final states containing one or more jets, which are challenging to probe at LHC due to large backgrounds, turn out to be promising at ILC due to clean environment. We would like to mention here that these results can be improved by using the multivariate analysis.

\begin{table}
 \begin{tabular}{|c|c|c|c|c|}
         \cline{2-5}
     \multicolumn{1}{c|}{}& \multicolumn{4}{|c|}{$\mathcal{L}_{5\sigma}$ (fb$^{-1}$)}\\  \cline{1-5}
   Benchmark points   & $2\ell+\mET$ & $1\ell+ 2j+\mET$ & $4\ell+\mET$ & $4j + \mET$ \\ \cline{1-5} 
  BP1 & 16 & 4 & 550 & 1.0\\ \hline
  BP2 & 23 & 3 & 150 & 0.9\\ \hline
  BP3 & 18 & 14 & 4500 & 1.8\\ \hline
  BP4 & 570 & 220 & $>10^4$ & 27\\ \hline
\hline
 \end{tabular}
 \caption{Required integrated luminosity for 5$\sigma$ significance reach based on cut-based analysis for the chosen benchmark points for $2\ell+\mET$, $1\ell + 2j+\mET$, $4\ell+\mET$, $4j + \mET$ respectively.}
 \label{compare}
 \end{table}

\section{Conclusion}
\label{conc}

In this work we study the signals for an $S_3$-symmetric 2HDM extended with two generations of VLLs. We impose an additional $Z_2$-symmetry through which the mixing between the SM fermions and VLLs is disallowed, since the SM fermions are even and the VLLs are odd under the aforementioned symmetry.  Thus the lightest neutral VLL turns out to be a viable DM candidate owing to the $Z_2$-symmetry.

We choose four representative benchmark points BP1, BP2, BP3 and BP4 corresponding to low, medium and high DM masses, which satisfy the constraints coming from vacuum stability, perturbative unitarity, electroweak precision variables and Higgs signal strength along with the DM constraints coming from relic density, direct and indirect detections. We have presented detailed collider analysis for four distinct channels to probe these BPs at 1 TeV ILC, namely $2\ell+\mET$, $1\ell+ 2j+\mET$, $4\ell+\mET$ and $4j+\mET$. 

For $2\ell+\mET$ and $1\ell+ 2j+\mET$ channels, we perform both cut-based and BDT analysis. All these channels originate from the pair production of the charged and neutral VLLs. In the $2\ell+\mET$ channel, both the $W^{\pm}$ decay leptonically and in the $1\ell+ 2j+\mET$ channel, one $W^{\pm}$ decays leptonically and the other hadronically and for $4j+\mET$ channel, both the $W^{\pm}$ decay hadronically. However, it is seen that one can probe $4j+\mET$ channel with 5$\sigma$ significance with integrated luminosity $< 2$ fb$^{-1}$ for the first 3 benchmark points and with 27 fb$^{-1}$ luminosity for BP4 even with the cut-based analysis. This is the best performing channel at 1 TeV ILC. $1\ell+ 2j+\mET$ channel at 1 TeV ILC is the second best performing channel as it requires only $\mathcal{O} (1)$ fb$^{-1}$ luminosity to probe the first three benchmark points and 8.6 fb$^{-1}$ luminosity to probe BP4 using MVA. The third well-performing channel is $2\ell+\mET$, where 5$\sigma$ significance is achieved for the first three BPs with luminosity $< 25$ fb$^{-1}$. However, due to small effective cross-section, one requires $216.3$ fb$^{-1}$ luminosity to attain 5$\sigma$ significance for BP4. The $4\ell+\mET$ channel performs the worst among all channels at ILC. Only for BP1 and BP2 with luminosity $550$ fb$^{-1}$ and 150 fb$^{-1}$ respectively, one can attain 5$\sigma$ significance. 
We find that a better sensitivity to heavier VLLs with high DM masses can be obtained at ILC in both the leptonic and hadronic 
channels, which proved more challenging and nearly impossible at HL-LHC due to smaller signal cross sections as well as 
large hadronic backgrounds. Thus ILC will prove to be a better hunting ground for such particles which have electroweak 
strength  interactions.


\section{Acknowledgement}
IC acknowledges support from DST, India, under grant number IFA18-PH214 (INSPIRE Faculty Award). NG and SKR would like to acknowledge support from the Department of Atomic Energy, Government of India, for the Regional Centre for Accelerator-based Particle Physics (RECAPP). 


\providecommand{\href}[2]{#2}\begingroup\raggedright\endgroup


\end{document}